\documentclass[prd,nofootinbib,preprint,superscriptaddress]{revtex4}
\pdfoutput=1
\usepackage[T1]{fontenc}
\usepackage[utf8x]{inputenc}
\usepackage{amsmath,amssymb}
\usepackage{epsfig}
\usepackage{graphicx}
\usepackage[usenames,dvipsnames]{color}
\usepackage{subfigure}
\usepackage{slashed}
\usepackage[colorlinks,citecolor=blue]{hyperref}
\usepackage{pdfpages}
\usepackage{color}
\usepackage{comment}

\DeclareUnicodeCharacter{2212}{\ensuremath{-}}

\def\bea{\begin{eqnarray}}
\def\eea{\end{eqnarray}}

\begin{document}
%\preprint{IP/BBSR/2015-4}
%\title{Imprints of ultralight PBH formed during phase transitions on gravitational waves}

%\title{Probing ultra-light PBH formation mechanism via unique doubly-peaked gravitational waves spectrum}

\title{Observable gravitational waves and $\Delta N_{\rm eff}$ with global lepton number symmetry and dark matter}

\author{Debasish Borah}
\email{dborah@iitg.ac.in}
\affiliation{Department of Physics, Indian Institute of Technology Guwahati, Assam 781039, India}

\author{Nayan Das}
\email{nayan.das@iitg.ac.in}
\affiliation{Department of Physics, Indian Institute of Technology
Guwahati, Assam 781039, India}

\author{Rishav Roshan}                             \email{r.roshan@soton.ac.uk}
\affiliation{School of Physics and Astronomy, University of Southampton, Southampton SO17 1BJ, United Kingdom}

\begin{abstract}
We study the possibility of testing a dark matter (DM) scenario embedded in a global lepton number symmetry $U(1)_L$ via gravitational waves (GW) and cosmic microwave background (CMB) observations. The spontaneous breaking of $U(1)_L$ symmetry generates the seesaw scale as well as DM mass dynamically. The (pseudo) Nambu-Goldstone boson, known as majoron, acquires non-zero mass due to soft symmetry breaking terms of quadratic type in the scalar potential, which eventually breaks $U(1)_L$ to its $Z_2$ subgroup. The spontaneous symmetry breaking, which effectively breaks $Z_2$, leads to the formation of domain walls (DW), posing a threat to successful cosmology, if allowed to dominate. As gravity does not respect any global symmetries, we consider higher dimensional operators suppressed by the scale of quantum gravity (QG) namely, $\Lambda_{\rm QG}$ which introduces the required bias leading to DW annihilation and emission of stochastic gravitational waves (GW) observable at near future experiments. The same operators also lead to decay of DM bringing interesting indirect detection aspects. While DM is produced non-thermally via scalar portal interactions, light majoron can give rise to additional $\Delta N_{\rm eff}$ within reach of future CMB experiments.
\end{abstract}

\maketitle

\section{Introduction}
Observations of non-zero neutrino mass and mixing, dark matter (DM) and baryon asymmetry of the Universe (BAU) \cite{ParticleDataGroup:2020ssz, Planck:2018vyg} have emerged as clear signs of beyond standard model (BSM) physics. While non-zero neutrino mass and mixing can be explained within the framework of canonical seesaw mechanisms like type-I seesaw \cite{Minkowski:1977sc, GellMann:1980vs, Mohapatra:1979ia,Sawada:1979dis,Yanagida:1980xy, Schechter:1980gr}, particle DM scenarios can be broadly categorised into thermal and non-thermal DM depending upon their production mechanisms. Among thermal and non-thermal DM scenarios, the weakly interacting massive particle (WIMP) and feebly interacting massive particle (FIMP), respectively, have been studied extensively in the literature. Recent reviews of WIMP and FIMP can be found in \cite{Arcadi:2017kky} and \cite{Bernal:2017kxu} respectively. Similarly, popular BSM scenarios to explain BAU can be categorized into baryogenesis \cite{Weinberg:1979bt, Kolb:1979qa} or leptogenesis \cite{Fukugita:1986hr}. Interestingly, leptogenesis can be realised within most of the seesaw frameworks where one or more of the heavy particles introduced for seesaw origin of neutrino mass can also be responsible for generating a non-zero lepton asymmetry to be converted into BAU later by electroweak sphalerons \cite{Kuzmin:1985mm}.

In spite of having several motivations for BSM physics mentioned above, none of the experiments across different frontiers have observed any convincing signs of it so far. Canonical seesaw models consistent with neutrino mass and leptogenesis typically have a very high scale for new physics, away from direct experimental reach. On the other hand, WIMP DM has not shown up in direct search experiments either \cite{LUX-ZEPLIN:2022qhg}. This might be indicative of the fact that DM is either feebly coupled or interact with the standard model (SM) particles via superheavy mediators. Such null results at DM search or other particle physics experiments searching for BSM physics have also motivated the search for complementary probes at other experiments. For example, some recent attempts have focused on probing high scale seesaw via stochastic gravitational wave (GW) observations \cite{Dror:2019syi, Blasi:2020wpy, Fornal:2020esl, Samanta:2020cdk, Barman:2022yos, Huang:2022vkf, Dasgupta:2022isg, Okada:2018xdh, Hasegawa:2019amx, Borah:2022cdx, Borah:2022vsu, Barman:2023fad, Borah:2023saq}. The particle physics scenarios explored in these works considered the presence of additional symmetries whose spontaneous breaking not only lead to the dynamical generation of seesaw scale but also leads to the generation of stochastic GW from topological defects like cosmic strings \cite{Dror:2019syi, Blasi:2020wpy, Fornal:2020esl, Samanta:2020cdk, Borah:2022vsu}, domain walls \cite{Barman:2022yos, Barman:2023fad,Saikawa:2017hiv,Roshan:2024qnv,Bhattacharya:2023kws,Blasi:2022ayo,Blasi:2023sej} or from bubbles generated at first order phase transition \cite{Huang:2022vkf, Dasgupta:2022isg, Okada:2018xdh, Hasegawa:2019amx, Borah:2022cdx, Borah:2023saq}. GW observations has also been used to probe even higher scales than seesaw, where quantum gravity (QG) effects can become important \cite{King:2023ayw, King:2023ztb, Borah:2022wdy}. In these works, the QG effects on explicit global discrete symmetry breaking was considered in the context of domain wall formation from spontaneous discrete symmetry breaking and consequent GW emission. The authors of \cite{King:2023ayw, King:2023ztb} also considered the impact of such QG effects on decay of DM which is stabilised by discrete symmetry at renormalisable level.

Motivated by these, we consider a global $U(1)$ symmetry related to the seesaw scale and DM stability and study QG effects leading to stochastic GW and DM decay. Connection to lepton sector and seesaw becomes natural if this Abelian global symmetry is identified as the global lepton number symmetry $U(1)_L$. In the simplest realisation of seesaw, the spontaneous breaking of $U(1)_L$ generates the right handed neutrino (RHN) masses or the seesaw scale dynamically. The same breaking also results in a massless Nambu-Goldstone (NG) boson, known as the majoron \cite{Chikashige:1980qk, Chikashige:1980ui, Gelmini:1980re,Manna:2022gwn}. Majoron can acquire a non-zero mass from explicit $U(1)_L$ breaking mass terms. When the singlet scalar responsible for breaking $U(1)_L$ spontaneously acquires a non-zero vacuum expectation value (VEV), it essentially breaks a $Z_2$ symmetry, remnant from $U(1)_L$ after explicit $U(1)_L$ breaking mass terms are included. Even without such explicit symmetry breaking terms, $U(1)_L$ is anomalous in the SM and broken to $Z_3$ by non-perturbative effects \cite{tHooft:1976rip, Lazarides:2018aev}\footnote{See \cite{Lazarides:2018aev, Brune:2022vzd} for non-minimal majoron models without domain walls.}. Spontaneous breaking of discrete symmetries like $Z_2$ leads to the formation of domain walls (DW) \cite{Zeldovich:1974uw, Kibble:1976sj, Vilenkin:1981zs,Saikawa:2017hiv,Roshan:2024qnv}, which can overclose the Universe, if allowed to dominate\footnote{Spontaneous breaking of global $U(1)_L$ can also lead to formation of topological defects known as cosmic strings \cite{Fu:2023nrn}. However, GW emission remains suppressed compared to particle emission from global strings \cite{Baeza-Ballesteros:2023say}.}. We then consider higher dimensional operators suppressed by QG scale as bias leading to disappearance of DW thereby generating stochastic GW signal detectable at ongoing and future experiments. We consider a Dirac fermion having chiral transformation under $U(1)_L$ to be the DM which remains accidentally stable even after $U(1)_L$ or $Z_2$ breaking, unlike earlier works \cite{King:2023ayw, King:2023ztb} where a separate $Z_2$ symmetry was introduced for DM stability. The $U(1)_L$ global symmetry therefore, generates the scale of seesaw, DM mass, ensures the stability of DM and also leads to DW formation in the presence of $Z_2$-preserving explicit $U(1)_L$ breaking terms generating majoron mass. The higher dimensional operators suppressed by QG scale can however, lead to DM decay opening up some indirect detection prospects. We discuss the details of DM which interacts with the SM via singlet scalar portal, and the effects of QG scale suppressed operators on GW emission from DW and on DM phenomenology. The RHNs not only generate light neutrino masses via seesaw mechanism, but can also generate the baryon asymmetry of the Universe via leptogenesis \cite{Fukugita:1986hr}. The emergence of $Z_2$ symmetry from a global $U(1)_L$, effects of QG on DW and DM decay, the presence of the pseudo NG boson or majoron bring complementary detection prospects at particle physics, gravitational waves as well as cosmology based experiments.

This paper is organised as follows. In section \ref{sec1}, we discuss the particle physics setup followed by the details of DM and majoron cosmology in section \ref{sec2}. In section \ref{sec3}, and section \ref{sec4} respectively, we discuss the details of DM decay and domain wall disappearance emitting gravitational waves due to QG scale suppressed operators. We finally conclude in section \ref{sec5}.

\section{The Setup}
\label{sec1}

\begin{table}[]
    \centering
    \begin{tabular}{|c|c|}
    \hline
     Fields    &  $SU(2)_L \times U(1)_Y \times U(1)_L$ charge\\
     \hline
     $ \chi_L $   & $(1, 0, Q_L) $ \\
     $ \chi_R $   & $(1, 0, Q_L+2) $ \\
     $ N_R $   & $(1, 0, 1) $ \\
     $ \ell_L $   & $(2, -\frac{1}{2}, 1) $ \\
    $ \ell_R $   & $(1, -1, 1) $ \\
     $ H $   & $(2, \frac{1}{2}, 0) $ \\
     $ \Phi $   & $(1, 0, -2) $ \\
         \hline
    \end{tabular}
    \caption{Relevant field content of the model.}
    \label{tab:content}
\end{table}
The relevant field content and respective transformations under $SU(2)_L \times U(1)_Y \times U(1)_L$ symmetry are shown in table \ref{tab:content} with the BSM fields being the right handed neutrino $N_R$, fermion DM $\chi$ and the singlet scalar $\Phi$ responsible for spontaneous symmetry breaking. While the specific choice of $Q_L$, the lepton number charge of $\chi_L$ does not affect our conclusion, we consider $Q_L \neq -1$, such that DM remains stable at renormalisable level. The relevant part of the Yukawa Lagrangian can be written as 
\begin{equation}
    -\mathcal{L}_Y \supset \frac{f_i}{2} \Phi \overline{N^c_{Ri}} N_{Ri} + Y_{\alpha i} \overline{\ell_{L_\alpha}} \tilde{H} N_{Ri} + y_\chi \overline{\chi_L} \Phi \chi_R + {\rm h.c.}
\end{equation}
The scalar potential is 
\begin{equation}\label{eq:potential}
    V(H, \Phi) = \lambda_H \left(H^\dagger H - \dfrac{v^2}{2}\right)^2 + \lambda_\Phi \left(|{\Phi}|^2 - \dfrac{v_{\Phi}^2}{2}\right)^2 +\lambda_{H \Phi} \left(H^\dagger H  - \dfrac{v^2}{2}\right)  \left(|{\Phi}|^2 - \dfrac{v_{\Phi}^2}{2}\right)\, 
\end{equation}
where $v (v_\Phi)$ denotes the VEV of $H (\Phi)$. The spontaneous breaking of $U(1)_L$ global symmetry after $\Phi$ acquires a VEV, leads to a massless Goldstone boson or majoron, identified as the angular part of the field $\Phi$. A non-zero majoron mass can be generated via explicit $U(1)_L$ breaking terms of the form 
\begin{equation}\label{eq:potential_1}
    V_{\slashed{L}} = - \frac{m^2}{4} (\Phi^2 + {\rm h.c.})
 \end{equation}
which breaks $U(1)_L$ to a remnant $Z_2$ symmetry in the scalar potential. It should be noted that we have considered the simplest form of explicit $U(1)_L$ breaking to generate the mass of majoron. Other forms of explicit $U(1)_L$ breaking will lead to different remnant symmetry. For example, explicit $U(1)_L$ breaking terms like $\Phi^3, \Phi^4$ will correspond to remnant symmetry $Z_3, Z_4$ respectively. Allowing, the explicit breaking to occur via non-renormalisable operators $\Phi^n/\Lambda^{n-4}$ will lead to bigger remnant symmetry groups like $Z_n$ where $n>4$. The details of the domain wall formation, evolution and emission of GW differs from the simplest $Z_2$ domain walls discussed in this work. Phenomenology of such $Z_n \, (n>2)$ domain walls can be found in \cite{Wu:2022stu, Wu:2022tpe, Bai:2023cqj} and references therein. 

After the spontaneous breaking of global $U(1)_{L}$ symmetry, we can represent $\Phi = (v_{\Phi} + \phi + i\eta)/\sqrt{2}$, where $\eta$ denotes the majoron. From the potential terms in Eq. \eqref{eq:potential} and Eq. \eqref{eq:potential_1}, the masses of Higgs ($h$), $\phi$ and majoron can be written as \cite{Gu:2010ys, Queiroz:2014yna}
\begin{align}
    m^2_{h} \simeq 2 \lambda_{H} v^2, \quad m^2_{\phi} = 2 \lambda_{\Phi} v^2_{\Phi}, \quad m^2_{\eta} \simeq m^2. 
\end{align}
The coupling between the majoron and heavy right-handed neutrino can be written as 
\begin{eqnarray}
    -\mathcal{L} \supset \frac{i f_{i}}{2\sqrt{2}} \eta  \overline{N_{i}} \gamma^{5} N_{i} = \frac{i m_{N_i}}{2 v_{\Phi}} \eta  \overline{N_{i}} \gamma^{5} N_{i}.
\end{eqnarray}
Similarly, the coupling between the majoron and DM can be written as
\begin{eqnarray}
    -\mathcal{L} \supset \frac{i y_{\chi}}{\sqrt{2}} \eta  \overline{\chi} \gamma^{5} \chi = \frac{i m_\chi}{v_{\Phi}} \eta  \overline{\chi} \gamma^{5} \chi.
\end{eqnarray}
The details of derivation of coupling between CP-odd and CP-even scalars with $N_{i} =N_{R_{i}} + N^{c}_{R_{i}}$  and $\chi = \chi_{L} + \chi_{R}$ are given in appendix \ref{appenA}.
Light neutrino mass arises from the type-I seesaw mechanism as
\begin{equation}
    m_\nu = -M_D m^{-1}_N M^T_D
\end{equation}
where $M_D = Y v/\sqrt{2}$ is the Dirac mass. Here $m_{N_{i}} = f_{i}v_{\Phi}/\sqrt{2}$, $m_{\phi}=\sqrt{2\lambda_{\Phi}}v_{\Phi}$, $m_{\chi}=y_{\chi}v_{\Phi}/\sqrt{2}$. The free parameters of the model, relevant for the phenomenology, are $f_{i}, Y_{\alpha i},  y_{\chi}, v_{\Phi}, \lambda_{\Phi},  \lambda_{H\Phi}$ and $m_{\eta}$. In terms of masses of RHNs, $\phi$ and $\chi$, the free parameters are $m_{N_{i}}, m_{\chi}, m_{\phi}, v_{\Phi},  Y_{\alpha i}, \lambda_{H\Phi}$ and $m_{\eta}$. We also denote the RHNs simply as $N$ hereafter.

Finally, there exists no exact (continuous or discrete) global symmetry in the theory of QG \cite{Kallosh:1995hi}. In other words, any global symmetry of a given effective field theory (EFT) is at best an approximate symmetry emergent in the IR \cite{Witten:2017hdv} and should be broken by a higher-dimensional operator. As a result, the remnant $Z_2$ present in this setup will also be explicitly broken by the higher dimensional operators \cite{Rai:1992xw} \footnote{See \cite{Beyer:2022ywc} and references therein for similar discussions in QCD axion models with Peccei-Quinn $U(1)$ global symmetry.}, 

\begin{eqnarray}
        \Delta V  &=& \frac{1}{\Lambda_{\rm QG}} (\alpha_{1}  \Phi^5 + \alpha_{2}  \Phi^3 H^2 + \alpha_{3}  \Phi H^4 ) \; , 
        \label{eq:bias}
    \end{eqnarray}
where $\Lambda_{\rm QG}$ denotes the scale of QG. Here we assume a common origin and therefore a common scale for the breaking of all global symmetries, and for simplicity, we take all the dimensionless coefficients in Eq. \eqref{eq:bias} to be of the same order, and we can make them of $\mathcal{O}(1)$ by redefining $\Lambda_{\rm QG}$.

\section{Cosmology of majoron and dark matter}
\label{sec2}
Majoron can have very interesting cosmology with the details being dependent upon its mass and interactions. Very light majorons in eV or sub-eV scale with negligible interactions can be approximated as a coherently oscillating scalar field, produced via the misalignment mechanism similar to axions \cite{Preskill:1982cy, Abbott:1982af, Dine:1982ah, Co:2019jts}. Depending upon the initial displacement or kinetic energy of the scalar field, majoron can play the role of cold dark matter in the Universe. In addition to that, majoron can also lead to spontaneous baryogenesis via leptogenesis \cite{Affleck:1984fy, Cohen:1987vi, Cohen:1988kt, Co:2019wyp, Chun:2023eqc, Ibe:2015nfa, Domcke:2020kcp, Co:2022aav, Berbig:2023uzs, Co:2020jtv, Chao:2023ojl, Barnes:2024jap, Barenboim:2024akt, Barenboim:2024xxa, Datta:2024xhg, Co:2024oek}.
In this work, we neither consider majoron DM nor spontaneous leptogenesis. This allows more available parameter space in singlet majoron model which otherwise gets constrained when majoron plays the role of DM \cite{Akita:2023qiz, deGiorgi:2023tvn}. Also, majoron DM is not perfectly stable but long-lived. However, one of our motivations is to study the effect of QG scale on lifetime of DM, which otherwise is perfectly stabilised by a global symmetry. Also considering thermal leptogenesis from RHN decay makes the scenario independent of any initial conditions. While leptogenesis can occur from CP violating out-of-equilibrium decays of right handed neutrinos in type-I seesaw, DM is in the form a separate fermion field which couples to the SM via radial and angular mode (majoron) of the singlet scalar field responsible for spontaneous breaking of global lepton number symmetry. We consider sizeable coupling of majoron with the SM bath such that it can be thermally produced. Such majorons can not only act like a portal between DM and SM but can also contribute to the effective relativistic degrees of freedom, observable at cosmic microwave background (CMB) experiments. Heavy majorons above the eV scale, which do not contribute to DM, are likely to decay into the SM particles without leaving any relic in the present Universe.

%\subsection{Themalisation of majoron}

\begin{figure}
    \centering
    \includegraphics[height=6.5 cm, width = 7cm]{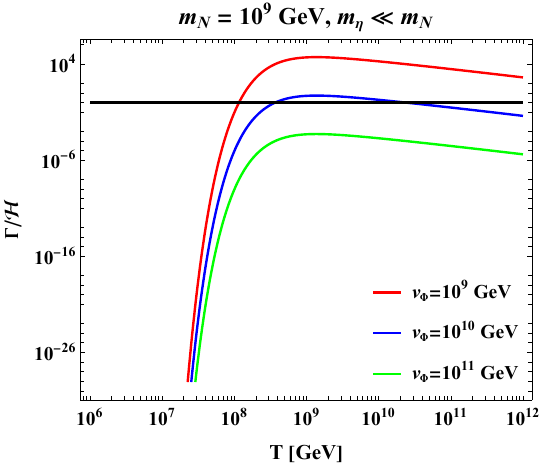}
    \includegraphics[height=6.1 cm, width = 7 cm]{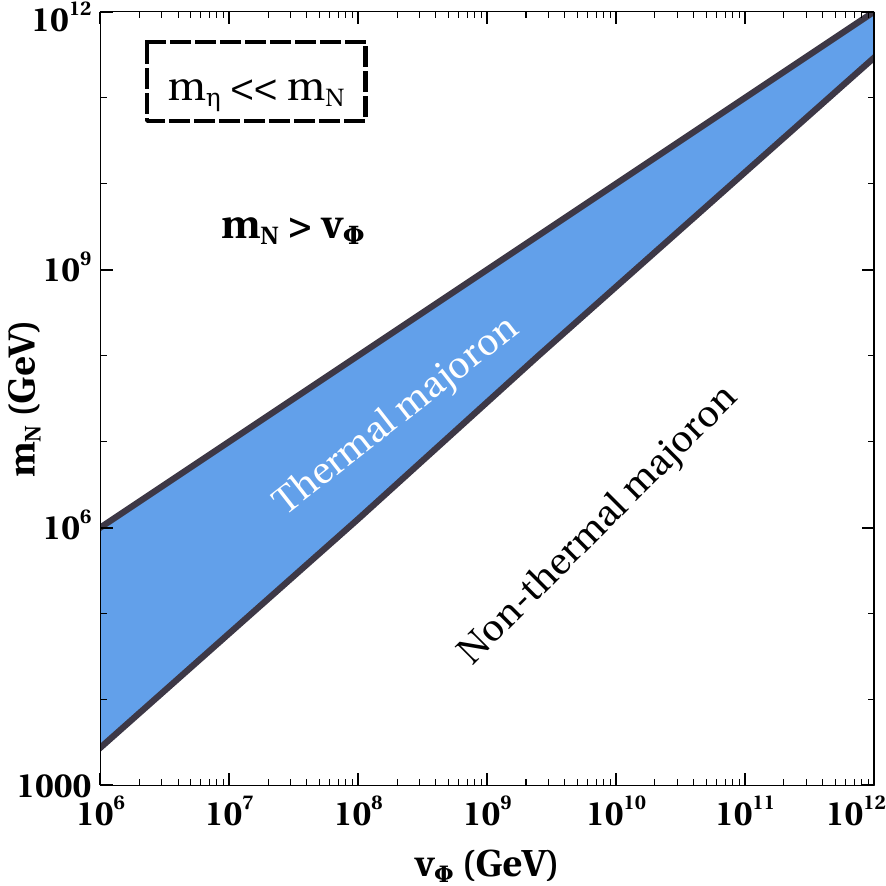}
    \caption{Left panel: majoron interaction rate in comparison to the Hubble expansion rate. Right panel: Parameter space in RHN mass $m_N$ and singlet scalar VEV $v_\Phi$ showing the thermalisation region of majoron for $m_{\eta} \ll m_{N}$.}
    \label{fig:themal_majoron}
\end{figure}

We first discuss the themalisation of majoron, $\eta$. For this, we have considered the scattering process, $N \eta \to N \eta$ and compare its rate with the Hubble expansion rate of the Universe $\mathcal{H}$. In the left panel plot of Fig. \ref{fig:themal_majoron}, we show the ratio of interaction rate ($\Gamma = n^{\rm eq}_{N}\langle\sigma v\rangle_{N\eta\to N\eta}$) to $\mathcal{H}$ as a function of temperature for three different values of $v_{\Phi}$.  For a larger $m_{N}/v_{\Phi}$ ratio (in other words, a larger $f$), we can have thermalised majoron. It is found that as long as $m_{\eta} \ll m_{N}$, the thermalisation of majoron is independent of its mass. In the right panel plot of Fig. \ref{fig:themal_majoron}, we show the parameter space in $m_{N}$ vs $v_{\Phi}$ plane where majoron thermalises. We have not considered the upper triangular region where $m_{N} > v_{\Phi}$, where the coupling can become non-perturbative.

Light majoron in sub-eV ballpark can contribute to the effective relativistic degrees of freedom $\Delta N_{\rm eff}$ tightly constrained by CMB observations as well as successful big bang nucleosynthesis (BBN). Majoron contribution to $\Delta N_{\rm eff}$ can be calculated based on the thermal history of majoron. The contribution to $\Delta N_{\rm eff}$ from light thermal majoron at the epoch of recombination is given as 
\begin{eqnarray}
    \Delta N_{\rm eff} \equiv \frac{\rho_{\eta} (T_{\rm CMB})}{\rho_{\nu, 1} (T_{\rm CMB})} = 0.027 \times \left(\frac{106.75}{g_{*}(T_{\rm dec})} \right)^{4/3}.
\end{eqnarray}
Here $\rho_{\eta}$ and $\rho_{\nu, 1}$ denotes the energy density of majoron and a single species of active neutrino respectively and $g_{*}$ denotes the effective degrees of freedom of SM particles. For all the blue shaded region shown in the right panel plot of Fig. \ref{fig:themal_majoron}, majoron gets decoupled at a temperature larger than $100$ GeV. This gives a constant contribution of thermal majoron $\Delta N_{\rm eff} =0.027$ for all $v_{\Phi}$. Future CMB experiment, specially, CMB-HD \cite{CMB-HD:2022bsz} can probe $\Delta N_{\rm eff}$ upto $0.014$ at $1\sigma$, keeping thermalised light majoron parameter space verifiable. We also check the non-thermal production of majorons from topological defects like cosmic strings and domain walls in our model \cite{Reig:2019sok} and find the corresponding abundance to be sub-dominant compared to DM, for the region of parameter space considered in our work.

DM couples to the SM bath or RHNs only via $\phi, \eta$ with a coupling proportional to $m_\chi/v_\Phi$. We consider $v_\Phi \geq 10^6$ GeV motivated from type-I seesaw, leptogenesis as well as domain wall related gravitational wave signatures to be discussed below. For such a high scale symmetry breaking, DM can not have large sizeable portal couplings required for thermal freeze-out, for typical thermal DM masses around the electroweak scale, below the unitarity limit \cite{Griest:1989wd}. Therefore, it is more natural to consider freeze-in dark matter \cite{Hall:2009bx}. Depending upon the mass spectrum of newly introduced particles, here we discuss two different scenarios for freeze-in DM.

\subsection{Dark matter with \boldmath{$m_{\phi}> 2 m_{N}$}} \label{sub1}
To begin with, we consider the mass spectrum as $m_{\phi}> 2 m_{N} > m_{\chi}> m_{\eta}$. We assume that RHNs are in thermal bath via the Yukawa interaction with SM leptons and Higgs. Similarly, $\phi$ is taken to be in thermal equilibrium initially. The DM is considered to be feebly interacting, produced through two possible ways : from decay of CP even scalar ($\phi \to \chi \Bar{\chi}$) and from annihilation of RHNs ($N\Bar{N} \to \chi\Bar{\chi}$ with $\eta$ and $\phi$ as the mediator).
Assuming $N$ to be in thermal equilibrium, the Boltzmann equations (BE) for $\phi$ and $\chi$ can be written as 
\begin{eqnarray} \label{BE1}
    \frac{d Y_{\phi}}{dx} &=& - \frac{\beta s}{\mathcal{H}x}\langle \sigma v \rangle _{\phi\phi \to N\Bar{N}} (Y^{2}_{\phi} - (Y^{\rm eq}_{\phi})^{2})  - 
    \frac{\beta}{\mathcal{H}x}\Gamma_{\phi\to N \Bar{N}} \frac{K_{1}(m_{\phi}/T)}{K_{2}(m_{\phi}/T)} (Y_{\phi}-Y^{\rm eq}_{\phi}) \\  &&  -\frac{\beta}{\mathcal{H}x}\Gamma_{\phi\to \chi \Bar{\chi}} \frac{K_{1}(m_{\phi}/T)}{K_{2}(m_{\phi}/T)} Y_{\phi}, \nonumber \\
    \frac{d Y_{\chi}}{dx} &=& \frac{\beta}{\mathcal{H}x} \Gamma_{\phi\to \chi \Bar{\chi}}\frac{K_{1}(m_{\phi}/T)}{K_{2}(m_{\phi}/T)} Y_{\phi} + \frac{\beta s}{\mathcal{H}x} \langle \sigma v \rangle _{N\Bar{N} \to \chi\Bar{\chi}} (Y^{\rm eq}_{N})^{2}.
\end{eqnarray}
Here, $Y$ and $s$ represent the comoving number density and entropy density respectively. The effect due to change of $g_{*,s}$ is incorporated into the parameter $\beta = \left[1 + \frac{T d g_{*,s}/dT}{3 g_{*,s}}\right]$ and $x=m_{0}/T$ where $m_{0}$ is some arbitrary reference mass. We consider $m_{0}=m_{\phi}$ for numerical calculations. $K_i$'s denote modified Bessel functions of the second kind whereas $\mathcal{H}$ denotes the Hubble expansion rate. The expression for relevant cross-sections are given in appendix \ref{appenB}.

\begin{figure}
    \centering
    \includegraphics[scale=0.85]{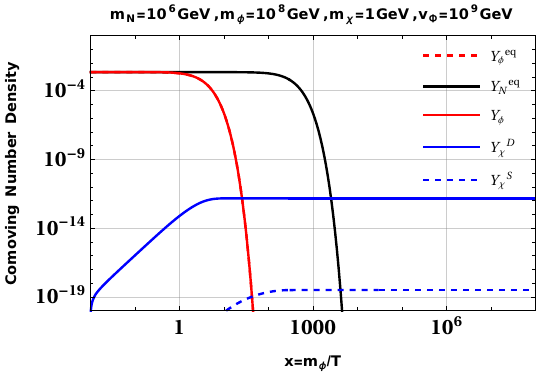}
    \includegraphics[scale=0.85]{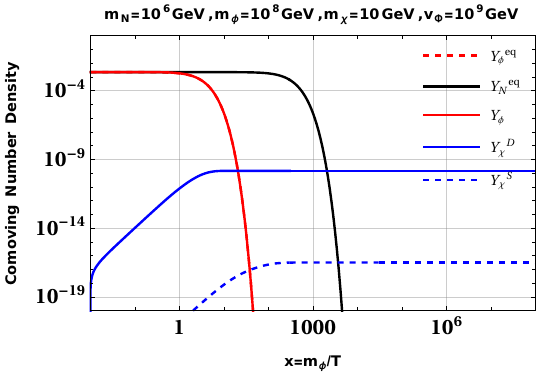}
    \caption{Evolution plot for $m_{\phi} > 2 m_{N}$ for two different $m_{\chi}=1$ GeV (left panel) and $m_{\chi}=10$ GeV (right panel). Due to the inverse decay of $N$, $\phi$ follows equilibrium abundance throughout its lifetime. The blue solid and dashed lines represent the respective contributions to DM production from the decay of $\phi$ and the annihilation of $N$.}
    \label{fig:evolution_t}
\end{figure}

In Fig. \ref{fig:evolution_t}, the evolution of $N$, $\phi$ and $\chi$ are shown. Here $Y^{D}_{\chi}$ and $Y^{S}_{\chi}$ denote the contribution to the DM production from the 1st and 2nd terms respectively in the BE for $Y_{\chi}$. Due to the inverse decay and annihilation of $N$, $\phi$ remains in equilibrium during its evolution. Also, the production of DM from annihilation is subdominant compared to the production from decay of $\phi$, as expected. Increasing DM mass from left to the right panel plots increases both $Y^{D}_{\chi}$ and $Y^{S}_{\chi}$ as both $\Gamma_{\phi\to \chi \Bar{\chi}}$ and $\langle \sigma v \rangle _{N\Bar{N}\to \chi \Bar{\chi}}$ are proportional to $m^{2}_{\chi}$.

\begin{figure}
    \centering
    \includegraphics[height=7cm, width=8cm]{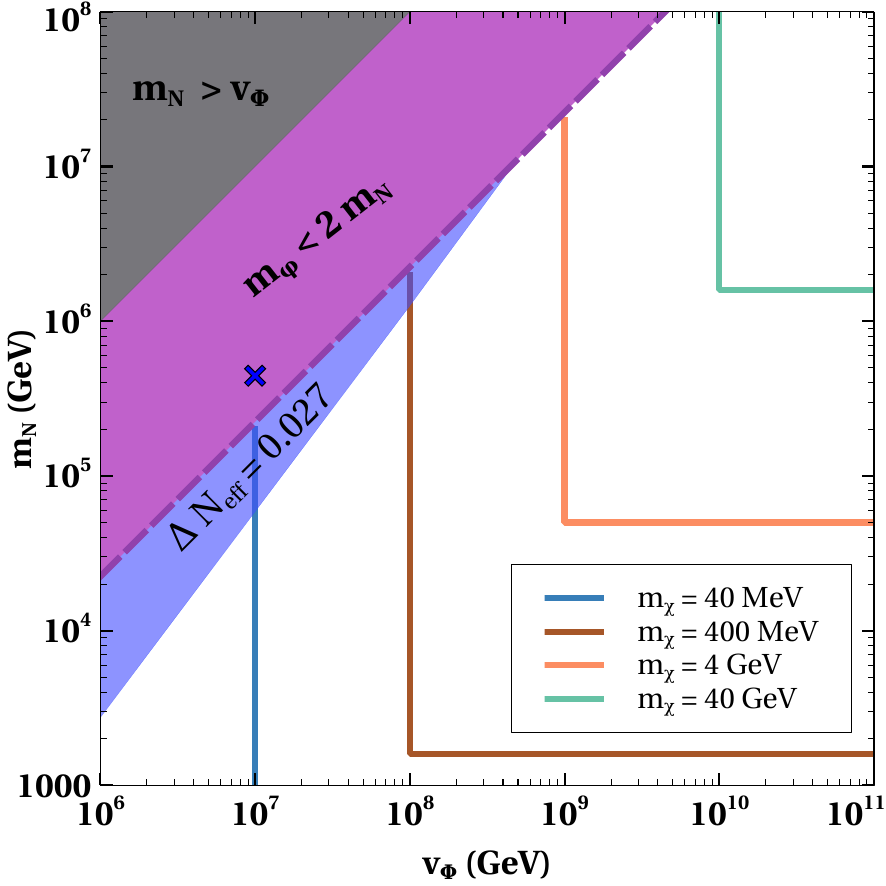}
    \caption{Parameter space in $v_{\Phi}$ vs $m_{N}$ plane with different colored contours denoting relic satisfying regions for different DM masses.}
    \label{fig:DM_eq}
\end{figure}

 Fig. \ref{fig:DM_eq} shows the parameter space in $v_{\Phi}$ vs $m_{N}$ plane with DM relic satisfying regions are shown as colored contours for different DM masses. Here, $\lambda_{\Phi}$ is taken as $10^{-3}$. This fixes $m_{\phi}$ for a given $v_{\Phi}$. In the gray and magenta region $m_{N} > v_{\Phi}$ and $m_{\phi} < 2 m_{N}$ respectively. The dashed magenta color line denotes $m_{\phi} = 2 m_{N}$. As depicted in the evolution plots of Fig. \ref{fig:evolution_t}, the production of dark matter from decay of $\phi$ dominates over the production from annihilation across the entire parameter space.  The four different colored contours correspond to different DM masses that satisfy correct relic abundance. Staring from the dashed magenta line, a change in $m_{N}$ does not change the DM relic for fixed $v_{\Phi}$ and $m_{\chi}$. This is because, in the vertical portion, the coupling between $N$ and $\phi$ (which is $\propto \frac{m_{N}}{v_{\Phi}}$) is larger enough so that $\phi$ remains in equilibrium due to the inverse decay of $N$ ($N\Bar{N}\to \phi$). However the situation reverses for a sufficiently small values of $m_{N}$ where $\phi$ does not always follow the equilibrium distribution. In that region, a different $v_{\Phi}$ (as well as $m_{\phi}$), does not alter DM abundances. Hence, the horizontal portion of the curves are obtained. Moreover, to indicate the parameter space that comes within the sensitivities of future CMB-HD experiment, we show the blue colored region  where majoron thermalises (from Fig. \ref{fig:themal_majoron}).

\subsection{Dark matter with \boldmath{$m_{\phi} < 2 m_{N}$}} \label{sub2}

\begin{figure}
    \centering
    \includegraphics{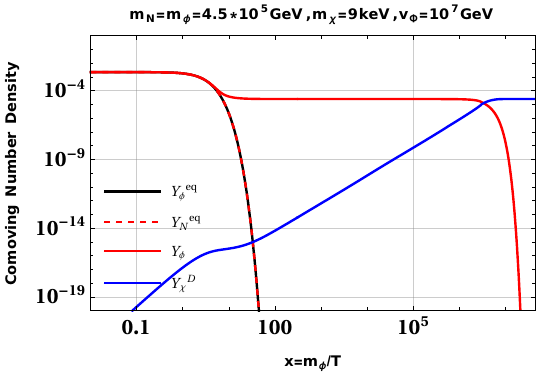}
    \caption{Evolution plot for $m_{\phi} < 2 m_{N}$. Here, we take $m_{N}=m_{\phi}$. The contribution from the $N\Bar{N} \to \eta \eta$ is negligible, hence not shown here.}
    \label{fig:evolution2}
\end{figure}

For the mass spectrum $2 m_{N} > m_{\phi} > m_{\chi}> m_{\eta}$, i.e. the magenta shaded region in Fig. \ref{fig:DM_eq}, the decay of $\phi \to N\Bar{N}$ ceases. Hence we have only the 1st and last term in the right hand side of equation \eqref{BE1}. Due to the presence of only one decay channel of $\phi$, the comoving DM abundance get enhanced. To see it, we consider a point in Fig. \ref{fig:DM_eq} denoted by $\times$. The evolution plot for this point is shown in Fig. \ref{fig:evolution2}. In the plot, $v_{\Phi}=10^{7}$ GeV and $\lambda_{\Phi}=10^{-3}$. This gives $m_{\phi}=4.5\times10^{5}$ GeV. Mass of $N$ is taken to be equal to $m_{\phi}$. As shown in the plot, for these choice of parameters a higher comoving abundance is obtained requiring DM mass to be as low as $9$ keV for correct relic density. Also due to such low DM mass, the decay width of $\phi$ is small ($\propto m^2_{\chi}$) resulting $\phi$ decaying very close to the BBN era. This brings two major problems namely, (i) the late decay of $\phi$ leads to a large the free-streaming length of DM (as shown in \cite{Decant:2021mhj, Biswas:2022vkq}) potentially giving hot DM that is already ruled out from structure formation constraint; (ii) DM mass range lies in the boundary of Tremaine-Gunn bound for fermionic dark matter \cite{PhysRevLett.42.407}. These problems get worse for a higher $v_{\Phi}$ making the scenario $m_{\phi}<2 m_{N}$ for $\lambda_{\phi}=10^{-3}$ disfavoured. Hence, for our current and future DM analysis, we only consider the region below the dashed magenta line in Fig. \ref{fig:DM_eq}.   

\subsection{Leptogenesis}
Apart from producing correct relic abundance, the current set up also allows to explain the observed baryon asymmetry of the Universe via thermal leptogenesis. The CP violating out-of-equilibrium decay of RHN neutrinos produce lepton asymmetry which is later converted to baryon asymmetry by SM sphalerons. Assuming hierarchical RHN mass spectrum, the baryon-to-photon ratio can be estimated as 
\begin{eqnarray} \label{eq:lep_asy}
    \eta_{B} \simeq 10^{-2} \epsilon_{1} \kappa_{1},
\end{eqnarray}
where $\epsilon_{1}$ denotes the CP asymmetry and $\kappa_{1}$ represents the efficiency factor. We use the maximum value of CP asymmetry parameter which is given as 
\begin{equation}\label{eq:epsilon}
    |\epsilon_{1}| \leq \frac{3 m_{N_{1}} \sqrt{(\Delta m_{\rm atm})^2}}{8 \pi v^2}.
\end{equation}
Here $(\Delta m_{\rm atm})^2 \simeq 2.4\times 10^{-3} \rm eV^{2}$ denotes the active neutrino atmospheric mass-squared difference. We consider efficiency factor to be $0.01$ that corresponds to strong wash-out regime. Apart from usual leptogenesis in type-I seesaw, in this setup we have extra annihilation channel, $N\Bar{N}\to \eta \eta$ that can keep RHNs in equilibrium for longer and hence dilute the final baryon asymmetry. However for strong washout regime and for equilibrium initial abundance of $N$, the effect from $N\Bar{N}\to \eta \eta$ can be neglected. With this, using Eq. \eqref{eq:lep_asy}, we get that the mass of RHN should be larger than $m_{N_{1}}\gtrsim 6\times 10^{10}$ GeV. Imposing the condition of correct DM
abundance along with correct baryon asymmetry, one needs $m_{\phi}> 2 m_{N}$ $\implies v_{\Phi} > \frac{2 m_{N}}{\sqrt{2 \lambda_{\Phi}}}$.  To make the calculations consistent with the DM and GW analysis, we consider $\lambda_{\Phi}=10^{-3}$ which gives
$v_{\Phi}$ larger than $ \sim 3\times 10^{12}$ GeV. Such large values of $v_{\Phi}$ are disfavoured from DW overclosure and DM properties, as we discuss below.  With the help of flavor leptogenesis, the minimum $m_{N_{1}}$ can be relaxed upto one order of magnitude \cite{Blanchet:2006be, Dev:2017trv} than the value obtained from Eq. \eqref{eq:lep_asy}.  Hence the corresponding value of $v_{\Phi}$ can also be reduced upto one order magnitude. For a even smaller value of $v_{\Phi}$, required amount of lepton asymmetry can be created via resonant leptogenesis \cite{Pilaftsis:2003gt}. Resonant leptogenesis can enhance CP asymmetry parameter upto $\mathcal{O}(0.1)$ where two RHNs have nearly the same mass, with the mass splitting being nearly equal to the its decay width. Therefore, correct baryon asymmetry is possible via resonant scenario where $v_{\Phi}$ is as small as $10^{6}$ GeV.

%%%%%%%%%%%%%%%%%%%%%%%%%%%%%%%%%%%%%%%%%%%%%%%%%%%%%%%%%%%%%%%
\section{Dark matter decay via quantum gravity effects}
\label{sec3}
In the present setup, the QG effects also break the $U(1)_L$ symmetry explicitly, and as a result, the DM decay can occur via dimension five operators of the type 
\begin{equation}
    -\mathcal{L}_{\rm decay} \supset \frac{Y_{D_\alpha}}{\Lambda_{\rm QG}} \overline{\ell_{L\alpha}}\tilde{H} \chi_R \Phi + {\rm h.c.},
    \label{DMdecay}
\end{equation}
where $\widetilde{H}=i\sigma_2 H^*$. As discussed earlier, we assume that all the global symmetries are broken by a common scale i.e. we simply consider all the dimensionless coefficients in Eq.~\eqref{eq:bias} and Eq. \eqref{DMdecay} to be of the same order, and we set them to be ${\cal O}(1)$ by redefining $\Lambda_{\rm QG}$. Looking at Eq. \eqref{DMdecay}, one notices that, once the electroweak symmetry is broken, the DM $\chi$ mixes with the SM neutrinos with a mixing angle \cite{Datta:2021elq,King:2023ztb},
\begin{equation}
\theta
 \simeq
 \sum_{\alpha=1,2,3}\left(\frac{Y_{D_\alpha} v_\Phi v}{\sqrt{2}\Lambda_{\text{QG}}m_\chi} \right)\; .   
 \label{eq:mixingangle}
\end{equation}

As a result of this mixing, the DM decays to the SM particles. This provides us a possibility to look for dark matter in the indirect search experiments. Once such example is constraining the DM decay life time by the detection of the galactic and extra-galactic diffuse X/$\gamma$-ray background. Due to the mixing, the DM can have two-body (one-loop level mediated via $W^\pm$ and SM charged leptons) and three-body (at tree-level mediated via $W^\pm$) decay channels that can dominate if kinematically allowed. At one-loop level $\chi$ can decay to photons and active neutrinos via $\chi \to \nu \gamma$, the decay rate of which is given by~\cite{Shrock:1982sc,Essig:2013goa}
\begin{equation}
\begin{split}
\tau_{\chi \rightarrow \nu \gamma} & \simeq\left(\frac{9 \alpha_{\mathrm{EM}} \sin ^2 \theta}{1024 \pi^4} G_F^2 m_\chi^5\right)^{-1} \\
& \simeq 1.8 \times 10^{17}~{\rm s}\left(\frac{10 \, \mathrm{MeV}}{m_\chi}\right)^5\left(\frac{\sin \theta}{10^{-8}}\right)^{-2} \; ,
\end{split}
\label{eq:lifetime}
\end{equation}
with $\alpha_{\rm EM} = 1/137$ being the electromagnetic fine-structure constant and $G_F$ denoting the Fermi constant. The three-body decay channel $\chi \to e^+ e^- \nu$, the decay rate of which can be approximately expressed as \cite{Ruchayskiy:2011aa,Essig:2013goa}
\begin{equation}
\begin{split}
    \tau_{\chi \to e^+ e^- \nu} & \simeq \left(   \frac{c_\alpha \sin^2 \theta}{96 \pi^3} G^2_F m^5_{\chi}\right)^{-1} \\
    & \simeq 2.4 \times 10^{15}~ {\rm s} \left(\frac{10 \mathrm{MeV}}{m_\chi}\right)^5\left(\frac{\sin \theta}{10^{-8}}\right)^{-2} \;, 
    \end{split}
    \label{eq:lifetime3}
\end{equation}
where $c_{\alpha}=(1+ 4 \sin^2 \theta_W + 8 \sin^4 \theta_W)/4 \simeq 0.59$ with $\theta_W$ being the weak mixing angle.
Contributions from the above two channels to the X/$\gamma$-ray fluxes are roughly at similar levels~\cite{Essig:2013goa}. The null detection of X-rays and $\gamma$-rays line sets a lower bound on the lifetime of the decaying DM that can further be converted into the constraints on the the DM mass and the mixing angle. The observed diffuse photon spectra data obtained from the HEAO-1~\cite{Gruber:1999yr}, INTEGRAL~\cite{Bouchet:2008rp}, COMPTEL~\cite{Sreekumar:1997yg} and EGRET~\cite{Strong:2003ey} satellites can restrict the parameter space of $m_\chi$ and $\theta$ within the range $0.01~{\rm MeV} \lesssim m_{\chi} \lesssim 100~{\rm MeV}$, which can be parameterized as $\theta^2 \lesssim 2.8 \times 10^{-18}({\rm MeV}/m_{\chi})^5$~\cite{Boyarsky:2009ix}. With  the help of the expression of $\theta$ in Eq.~\eqref{eq:mixingangle}, we have
\begin{equation}
\left(\frac{3 v_\Phi v}{\sqrt{2}\Lambda_{\text{QG}}m_\chi}\right)^2 \lesssim 2.8\times10^{-18}\left(\frac{\text{MeV}}{m_\chi}\right)^5 \; ,
\label{DM-decay}
\end{equation}
where we consider that the DM interacts with different generations of neutrinos with the identical ${\cal O}(1)$  strength. On top of this, the Fermi Large Area Telescope (Fermi-LAT)~\cite{Fermi-LAT:2012pls,Fermi-LAT:2012ugx} has presented a dedicated line search of the diffuse $\gamma$-ray background, the null result of which also constrains the lifetime of the DM within the mass range $1~{\rm GeV} \lesssim m_\chi \lesssim 1~{\rm TeV}$.  Here we take the 95\% C.L. limit on the lifetime of decaying DM given by ref.~\cite{Fermi-LAT:2015kyq}, which assumes a Navarro-Frenk-White (NFW) profile for the DM distribution~\cite{Navarro:1995iw}.

DM decay also faces stringent constraints from CMB, if the decay occurs after or during the recombination. The  DM decay can re-ionize the intergalactic medium, and hence can modify the CMB power spectrum.  Accurate measurements of the CMB spectrum have been implemented by recent experiments including WMAP~\cite{WMAP:2012nax}, ACT~\cite{ACTPol:2014pbf}, SPT~\cite{Hou:2012xq} and Planck~\cite{Planck:2018vyg}. The 95\% C.L. lower bounds on the DM decay lifetime have been given in ref.~\cite{Slatyer:2016qyl}, showing that $\tau_{\rm DM} \gtrsim 10^{25}~{\rm s}$ for decays into both $e^+e^-$ pairs and photons. These DM decays can also introduce radio signals originating inside the DM-dominated galaxies and clusters, if the produced $e^+e^-$ pairs undergo energy loss via electromagnetic interactions in the interstellar medium. Such radio waves can be tested by radio telescopes like the Square Kilometer Array (SKA) radio telescope~\cite{Colafrancesco:2015ola}. It is found that the DM decay width up to $\Gamma_\text{DM}\gtrsim10^{-30}~{\rm s}^{-1}$ is detectable at SKA assuming  100-hour observation time~\cite{Dutta:2022wuc}. We will incorporate these constraints while showing the final allowed parameter space, to be discussed below.

%%%%%%%%%%%%%%%%%%%%%%%%%%%%%%%%%%%%%%%%%%%%%%%%%%%%%%%%%%%%%%%

\section{Domain walls and Gravitational Waves}
\label{sec4}
Domain walls, a cosmological catastrophe, are 2-dimensional topological defects resulting from a spontaneous symmetry breaking  (SSB) of a discrete symmetry like $Z_2$, as also considered in the present setup. Once formed, their energy density varies inversely with the scale factor ($a^{-1}$) which is much slower than that of matter ($a^{-3}$) or radiation ($a^{-4}$). As a result, they can dominate the energy budget of the Universe at some later stage and can alter the fate of CMB observations or successful light nuclei synthesis during BBN. These cosmological catastrophes can be avoided by introducing an energy bias in the scalar potential that makes the vacuum unstable. The above scaling of the DW is true assuming the radiation-dominated era when the DWs were relativistic. At a later stage, the DW dynamic is dominated by its tension force that stretches the DWs up to the horizon size. Numerical studies \cite{Press:1989yh, Garagounis:2002kt, Oliveira:2004he, Leite:2011sc, Leite:2012vn} have shown that the evolution of DW in this regime can be described by the scaling solution. Here, their energy density evolves according to the simple scaling law $\rho_{\rm DW}\propto t^{-1}$ where $\rho_{\rm DW}$ denotes the energy density of the DW and their typical size is given by Hubble radius $\sim t^2$. In the scaling regime, their energy density can be expressed as,
\begin{equation}
    \rho_{\rm DW}=\sigma\frac{\mathcal{A}}{t},
\end{equation}
with $\mathcal{A}\simeq 0.8$ being the area parameter \cite{Hiramatsu:2013qaa} and $\sigma$ being the surface tension of the DW. In the present setup this can be achieved with the help of QG effect as it is expected to break any global symmetry at high scale. Following this, one can write the energy bias after replacing fields with their VEVs, this leads to a bias contribution to the effective potential of the form
\begin{equation}
    V^{}_{\rm bias} \simeq \frac{1}{\Lambda^{}_{\rm QG}}\left( v^5_\Phi + \frac{v^3_\Phi v^2}{2} + \frac{v^{}_\Phi v^4}{4} \right) \; .
    \label{eq:model-bias}
\end{equation}

Due to this uplifting of degeneracy, the DW starts to collapse and annihilate, resulting in the production of SGWB whose signals can be tested in the present and future GW detectors. Under the assumption that $v_\Phi \gg v$, a sufficiently large $V^{}_{\rm bias}$ can be achieved rendering the production of observable GW signals.  Note that to have a DW, a very large $ V^{}_{\rm bias}$ must be avoided according to the prediction of percolation theory \cite{ Saikawa:2017hiv}. Under such a hierarchy, the bias potential is dominated by,
\begin{equation}
    V^{}_{\rm bias} \simeq \frac{v^5_\Phi}{\Lambda^{}_{\rm QG}}  \; .
    \label{eq:model-bias_new}
\end{equation}
Before delving into the details of GW production from the annihilation of the DW, we first like to discuss the constraints that appear on the $V^{}_{\rm bias}$ as a result of DW annihilations to the SM particles. If the energy bias of the DW is very small, they live for a very long time ($t_{\rm ann}\propto \frac{1}{V_{\rm bias}}$, with $t_{\rm ann}$ being the annihilation time of the DW). Requiring that their collapse happens before they dominate the energy budget of the Universe puts a lower bound on the magnitude of $V_{\rm bias}$ \cite{ Saikawa:2017hiv},
\begin{equation}
V_{\rm bias}^{1/4}>2.18\times10^{-5} {\rm GeV}~ C_{\rm ann}^{1/4}\mathcal{A}^{1/2}\left(\frac{\sigma^{1/3}}{10^3\rm GeV}\right)^{3/2},
\label{vbias_LB}
\end{equation}
where $C_{\rm ann}\simeq 2$ is a dimensionless constant, and $\sigma=\sqrt{\frac{8\lambda_\Phi}{9}}v_\Phi^3$ is the surface tension of the DW. Even if Eq. \eqref{vbias_LB} is satisfied and the DW are annihilated away before their dominance, their decay products (assuming they are the SM particles) can still destroy the light elements created at BBN epoch. This demands $t_{\rm ann}\leq 0.01$ s and hence an additional constraint on the magnitude of the energy bias can be obtained as,
\begin{equation}
V_{\rm bias}^{1/4}>5.07\times10^{-4} {\rm GeV}~ C_{\rm ann}^{1/4}\mathcal{A}^{1/4}\left(\frac{\sigma^{1/3}}{10^3\rm GeV}\right)^{3/4}.
\label{vbias_LB2}
\end{equation}

The production of the GW from DW annihilation has been investigated in detail, for example, see refs. \cite{Vilenkin:1981zs, Gelmini:1988sf, Larsson:1996sp, Hiramatsu:2013qaa, Hiramatsu:2012sc, Saikawa:2017hiv,Roshan:2024qnv,Bhattacharya:2023kws}.  Assuming that the DW annihilate instantaneously ($t=t_{\text{ann}}$) during the radiation-dominated era, the peak frequency $f_{p}$ and peak energy density spectrum $\Omega_{p}h^2$ of GW at present can be expressed as \cite{Saikawa:2017hiv, Chen:2020wvu}
 \begin{align}\label{fpeak}
     f_{p}&\simeq 3.75\times10^{-9}~\text{Hz}\times C_{\rm ann}^{-1/2}\mathcal{A}^{-1/2}\bigg(\frac{10^3~\text{GeV}}{\mathcal{\sigma}^{1/3}}\bigg)^{3/2}\bigg(\frac{V_{\text{bias}}^{1/4}}{10^{-3}~\text{GeV}}\bigg)^{2}\,,\nonumber\\
     \Omega_{p}h^2&\simeq 5.3\times10^{-20}\times\tilde{\epsilon}_{\text{GW}}~C_{\rm ann}^{2}\mathcal{A}^{4}\bigg(\frac{\sigma^{1/3}}{10^3~\text{GeV}}\bigg)^{12}\bigg(\frac{10^{-3}~\text{GeV}}{V_{\text{bias}}^{1/4}}\bigg)^{8},
 \end{align}
 where $\tilde{\epsilon}_{\text{GW}}\simeq0.7$~\cite{Hiramatsu:2013qaa} denotes the fraction of energy radiated into GW and can be regarded as a constant in the scaling regime. The typical feature of the GW spectrum from DW is that it follows a broken power law, where the
breaking point has a frequency determined by the annihilation time and the peak amplitude
is determined by the energy density in the domain walls as can be seen from Eq. \eqref{fpeak}. To depict the GW spectrum, we adopt the following parametrization for a broken power-law spectrum~\cite{Caprini:2019egz, NANOGrav:2023hvm}
\begin{eqnarray}
 \Omega_{\rm GW}h^2_{} = \Omega_p h^2 \frac{(a+b)^c}{\left(b x^{-a / c}+a x^{b / c}\right)^c} \ ,
\label{eq:spec-par}
\end{eqnarray}
where $x \equiv f/f_p$, and $a$, $b$ and $c$ are real and positive parameters. Here the low-frequency slope $a = 3$ can be fixed by causality, while numerical simulations suggest $b \simeq c \simeq 1$~\cite{Hiramatsu:2013qaa}.

\begin{figure}[htb!]
  \centering
  \includegraphics[scale=0.4]{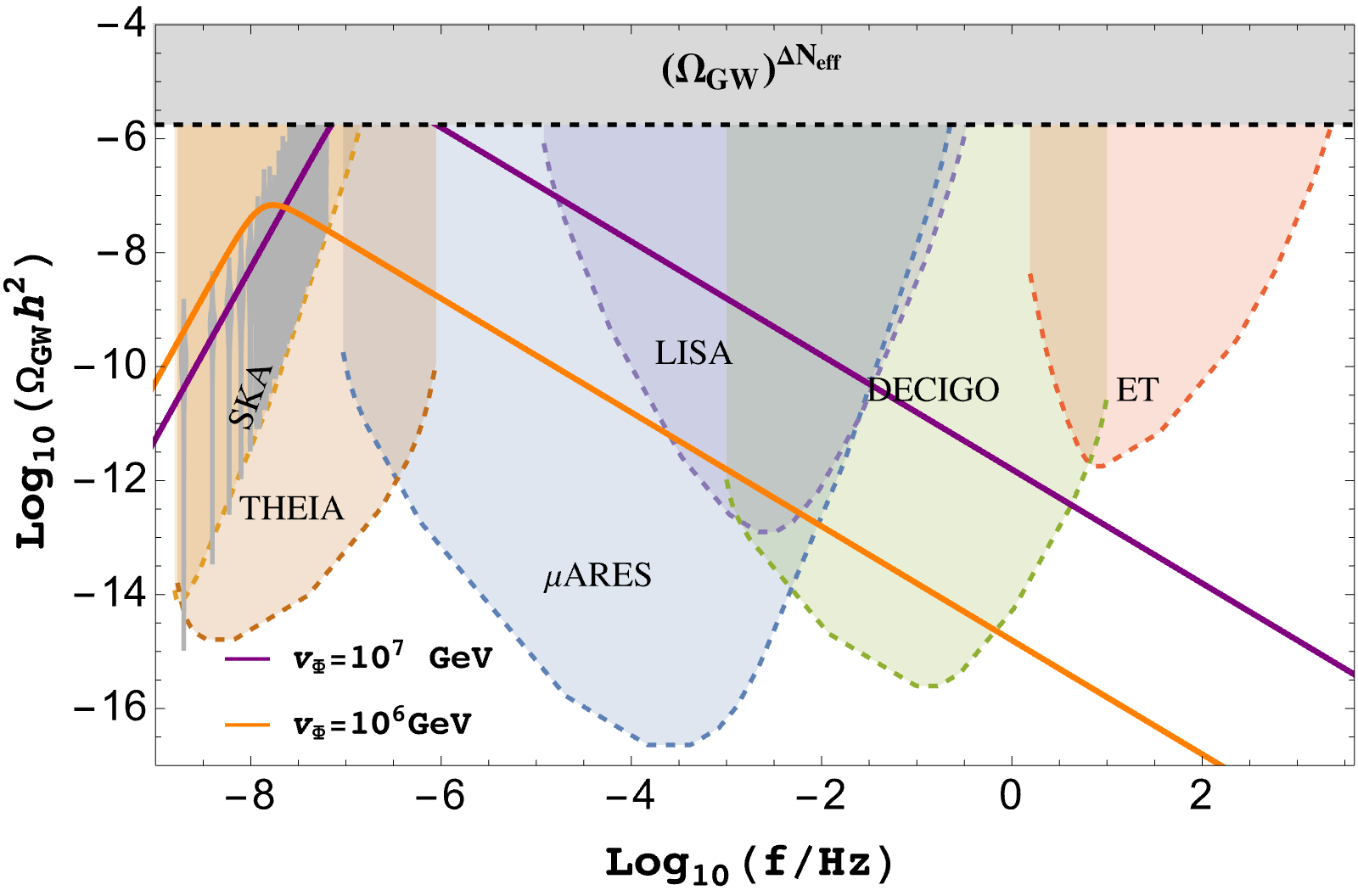} 
  \includegraphics[scale=0.4]{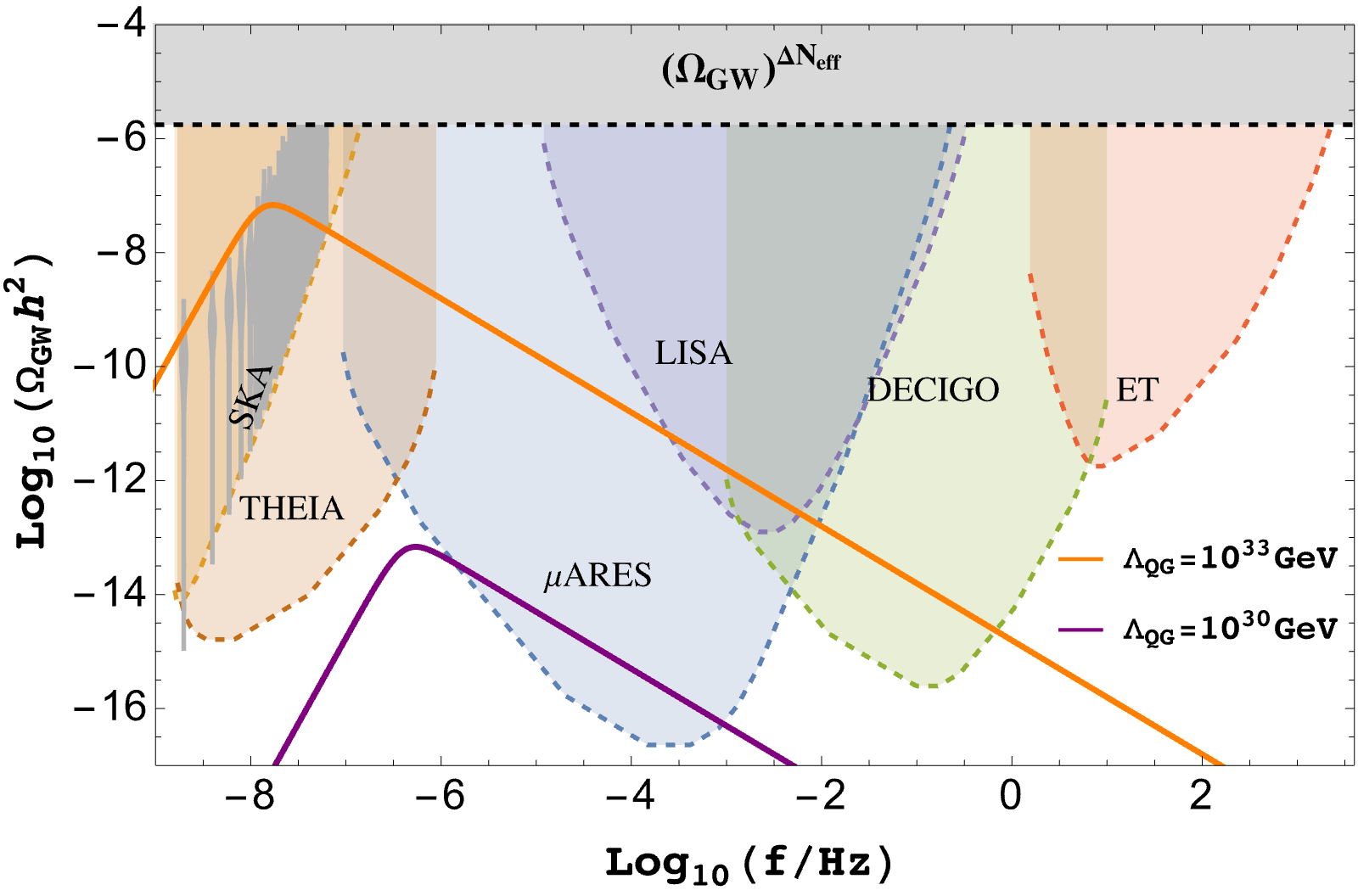}
\caption{The orange and purple curve show the GW spectrum generated from the annihilation of the DW. In both the panels, we set $\lambda_{\Phi}=10^{-3}$ just for the demonstrative purpose.  While in the left Panel we fix $\Lambda_{\rm QG}=10^{33}~\rm{GeV}$ and vary $v_\Phi$, in the right panel we fix $v_\Phi=10^6~\rm{GeV}$ and vary $\Lambda_{\rm QG}$. The gray color region denotes the portion excluded from $\Delta N_{\rm eff}$ bound. }
  \label{fig:DW_spec1}
\end{figure}

Apart from the constraint on DW from equations \eqref{vbias_LB} and \eqref{vbias_LB2}, a large amplitude of GW from DW annihilation can also contribute to extra effective number of relativistic species, $\Delta N_{\rm eff}$. This gives an upper bound on the GW energy denisty. Considering the Planck 2018 results on $\Delta N_{\rm eff}$ at the time of BBN, the constraint on GW energy density is given as \cite{Domenech:2020ssp}
\begin{eqnarray}
    \Omega_{\rm GW , BBN} < \frac{7}{8} \left(\frac{4}{11}\right)^{4/3} \Delta N_{\rm eff} \sim 0.05.
\end{eqnarray}
Hence, the maximum energy density of GW today which is allowed from $\Delta N_{\rm eff}$ bound can be written as 
\begin{eqnarray}
    \Omega_{\rm GW}^{\rm \Delta N_{\rm eff}}h^2 = 0.39 \left(\frac{g_{*}(T_{\rm BBN})}{106.75}\right)^{-1/3} \Omega_{r,0}h^2 \Omega_{\rm GW , BBN}  \simeq 1.75\times 10^{-6},
\end{eqnarray}
where $\Omega_{r,0}h^2\sim 4.18\times10^{-5}$ represents current radiation energy density fraction. Therefore, to be consistent with this bound, the peak energy density of GW from DW, $\Omega_{p}h^2$, should be less than $ \Omega_{\rm GW}^{\rm \Delta N_{\rm eff}}h^2$.

In Fig. \ref{fig:DW_spec1}, we present the GW spectrum for different choices of model parameters.  For both the plots we set $\lambda_\Phi=10^{-3}$.  In Fig. \ref{fig:DW_spec1}, the gray violins indicate the GW spectrum observed by the NANOGrav. We notice that the spectra shown by the orange color lines in the Fig. \ref{fig:DW_spec1} are in good agreement with the NANOGrav 15-year result. Such domain wall interpretation of the NANOGrav 15-year data was also discussed in earlier works, see \cite{Barman:2023fad} and references therein. Using Eq.~\eqref{eq:spec-par}, we also evaluate the sensitivity curves of the future GW detectors ET~\cite{Punturo:2010zz}, LISA~\cite{LISA:2017pwj}, DECIGO~\cite{Kawamura:2020pcg}, $\mu$Ares~\cite{Sesana:2019vho}, SKA~\cite{Janssen:2014dka}, IPTA \cite{Hobbs_2010} and THEIA~\cite{Garcia-Bellido:2021zgu} by calculating the signal-to-noise ratio (SNR)~\cite{Maggiore:1999vm,Allen:1997ad}
\begin{equation}
    \varrho=\left[n_{\mathrm{det}} t_{\mathrm{obs}} \int_{f_{\min }}^{f_{\max }} d f\left(\frac{\Omega_{\text {signal }}(f)}{\Omega_{\text {noise }}(f)}\right)^2\right]^{1 / 2} \; ,
    \label{eq:SNR}
\end{equation}
where $n_{\rm det} = 1$ for auto-correlated detectors and  $n_{\rm det} = 2$ for cross-correlated detectors, $t_{\rm obs}$ denotes the observational time, and $\Omega_{\text {noise }}$ represents the noise spectrum expressed in terms of the GW energy density spectrum~\cite{Schmitz:2020syl}. Integrating $(\Omega_{\rm signal}/\Omega_{\rm noise})^2$ over the sensitive frequency range of individual GW detectors, we obtain the SNRs for the GW spectra. Now, in the left panel, we fix the QG scale at $\Lambda_{\rm QG}=10^{33}$ GeV while we vary $v_\Phi$. Following Eq. \eqref{fpeak}, one finds that for dimension 5 operator, $\Omega_{p}h^2\propto (\lambda_\Phi^2 \Lambda_{\rm QG}^2)/v_\Phi^2$ while $f_p\propto v_\Phi/(\lambda_\Phi^{1/4}\Lambda_{\rm QG}^{1/2})$. As expected, a  larger $v_\Phi$ results in a smaller $\Omega_ph^2$ and a larger peak frequency. On the other hand, a larger $\Lambda_{\rm QG}$ scale corresponds to a larger  $\Omega_ph^2$ but a smaller peak frequency, this behavior is visible in the right panel of Fig. \ref{fig:DW_spec1}.

\begin{table}[!htb]
\centering
\begin{tabular}{c c c c c c c c}
\hline \hline 
BPs & $\Lambda_{\rm QG}$  (GeV)& $v_\Phi$ (GeV) & $m_{N}$ (GeV)& $m_{\chi}$ (MeV)& $\Delta N_{\rm eff}$ \\
 \hline 
 BP1 & $10^{33}$ & $10^7$ & $10^{5}$ & 40 & 0.027\\
 BP2 &$10^{33}$ & $10^6$ & $10^{4}$ & 4 & 0.027 \\
 \hline \hline
\end{tabular}
\caption{Set of benchmark values of $\Lambda_{\rm QG}$ and $v_{\Phi}$.}
\label{tab:bp}
\end{table}

\begin{figure}[htb!]
  \centering
  \includegraphics[scale=0.4]{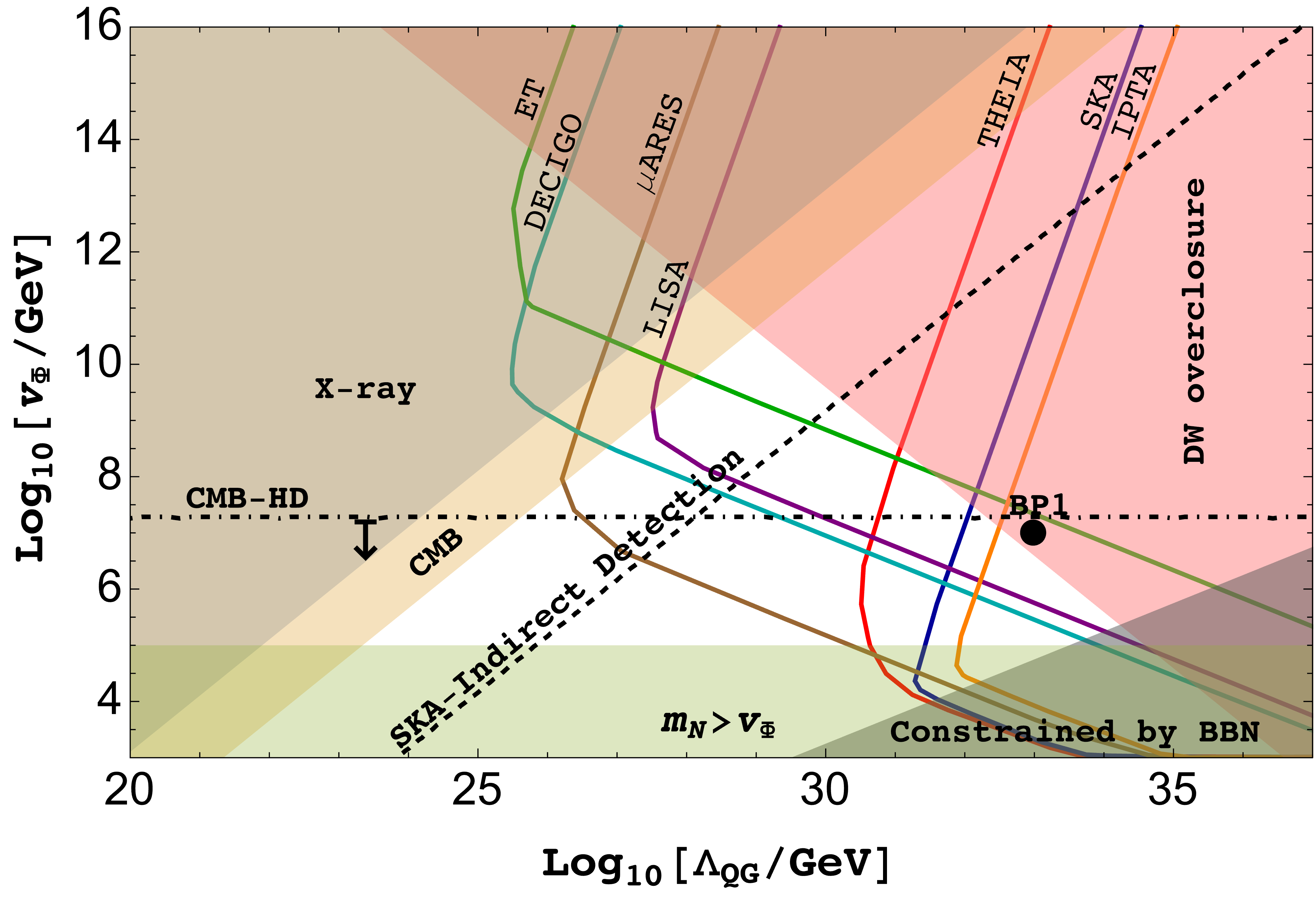}
  \includegraphics[scale=0.4]{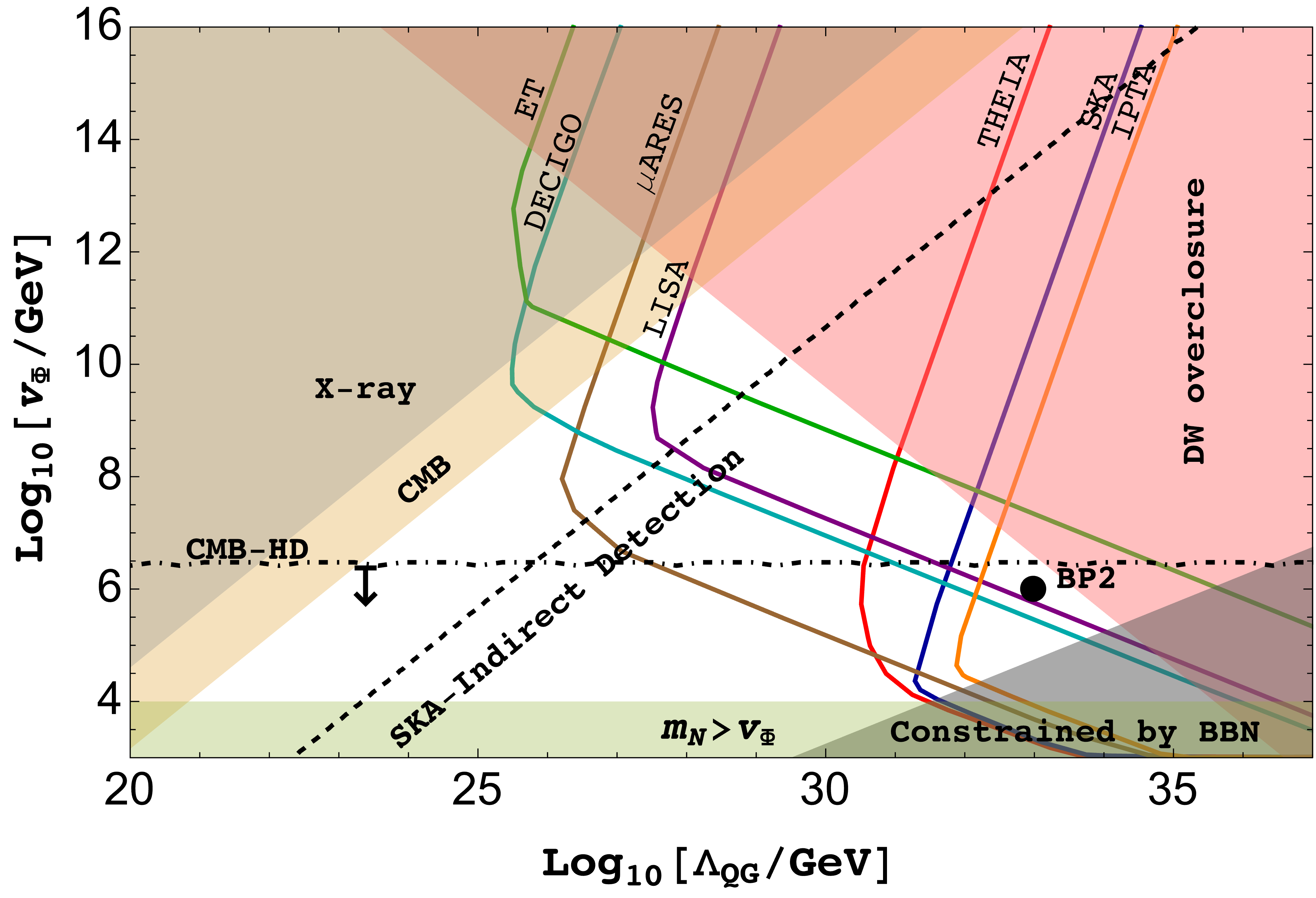}
\caption{Combined constraints on $\Lambda_{\rm QG}$ and $v_\Phi$ from indirect DM detection and GW observations with $\lambda_{\Phi}=10^{-3}$. The light gray and yellow-shaded regions denote the excluded regions by X-ray and CMB observations respectively. The black dashed lines label the testing capabilities of the upcoming SKA telescope. The black dots correspond to the BPs shown in Table \ref{tab:bp}. The red, blue, brown, cyan, purple, green, and orange curves correspond to the sensitivities of THEIA, SKA, $\mu$Ares, DECIGO, LISA, ET, and IPTA with SNR = 10. The gray-shaded regions are excluded by the requirement of BBN. The pink-shaded region corresponds to the scenario where DW may overclose the Universe at an early epoch. The region below the horizontal dot-dashed line is within reach of CMB-HD experiment.}
  \label{fig:summary}
\end{figure}

In Fig. \ref{fig:summary}, we summarise our result. For the demonstrative purpose, we consider different benchmark points (BPs) shown in Table \ref{tab:bp}. While in the left panel, we show BP1, in the right panel, we show BP2. In both plots, the light gray and the yellow-shaded regions are excluded from the null detection of X-ray signals originating from the DM decay and CMB observations respectively, as discussed in the previous section. The black dashed lines label the testing capabilities of the upcoming SKA telescope. The pink and gray-shaded regions are excluded from the constraints coming from the DW overclosure and BBN. The BBN limit arises from the requirement of not disrupting light nuclei abundance from DW annihilations at late epochs. We would also like to point out that the $\Delta N_{\rm eff}$ bounds on the GW is much weaker than the one obtained from the DW overclosure and hence we do not show it in the plots. The parameter space below the horizontal dot-dashed line remains within the sensitivity of future CMB experiment CMB-HD \cite{CMB-HD:2022bsz} which can measure the enhanced $\Delta N_{\rm eff}$ due to light majoron. Next, we mark the threshold SNR $\varrho = 10$ for different detectors by the solid-colored curves, which shows the parameter ranges detectable by these GW detectors. From Fig. \ref{fig:summary} and Fig. \ref{fig:DW_spec1}, it is clear that the BPs considered lie within the detectable ranges of almost all the GW detectors except ET. Note that even though the BP1 is allowed by the constraints that come from the indirect searches of the DM and also lies within the reach of different GW experiments, it still gets disallowed as it falls in the region of parameter space where the DW can overclose the Universe. Finally, the predictability of the model is also enhanced as one of the BPs considered in Table \ref{tab:bp} is already disfavored while other parameter space remain within reach of GW, CMB and DM indirect detection experiments. As can be seen from the summary plots shown in Fig. \ref{fig:summary}, the successful DM, DW and GW phenomenology requires the QG scale to be $\Lambda_{\rm QG} \gtrsim 10^{21}$ GeV. The scale needs to be even higher for the GW emitted by DW annihilation to remain within experimental sensitivities. Such large QG scale can be natural in theories where the global symmetries are broken by non-perturbative instanton effects for example, D-brane in string theory \cite{Blumenhagen:2006xt,Florea:2006si,Blumenhagen:2009qh} or gravitational instantons \cite{Giddings:1987cg,Lee:1988ge,Abbott:1989jw,Coleman:1989zu}.

\subsection{Analysis with dimension 7 operator}
\begin{figure}[htb!]
  \centering
  \includegraphics[scale=0.4]{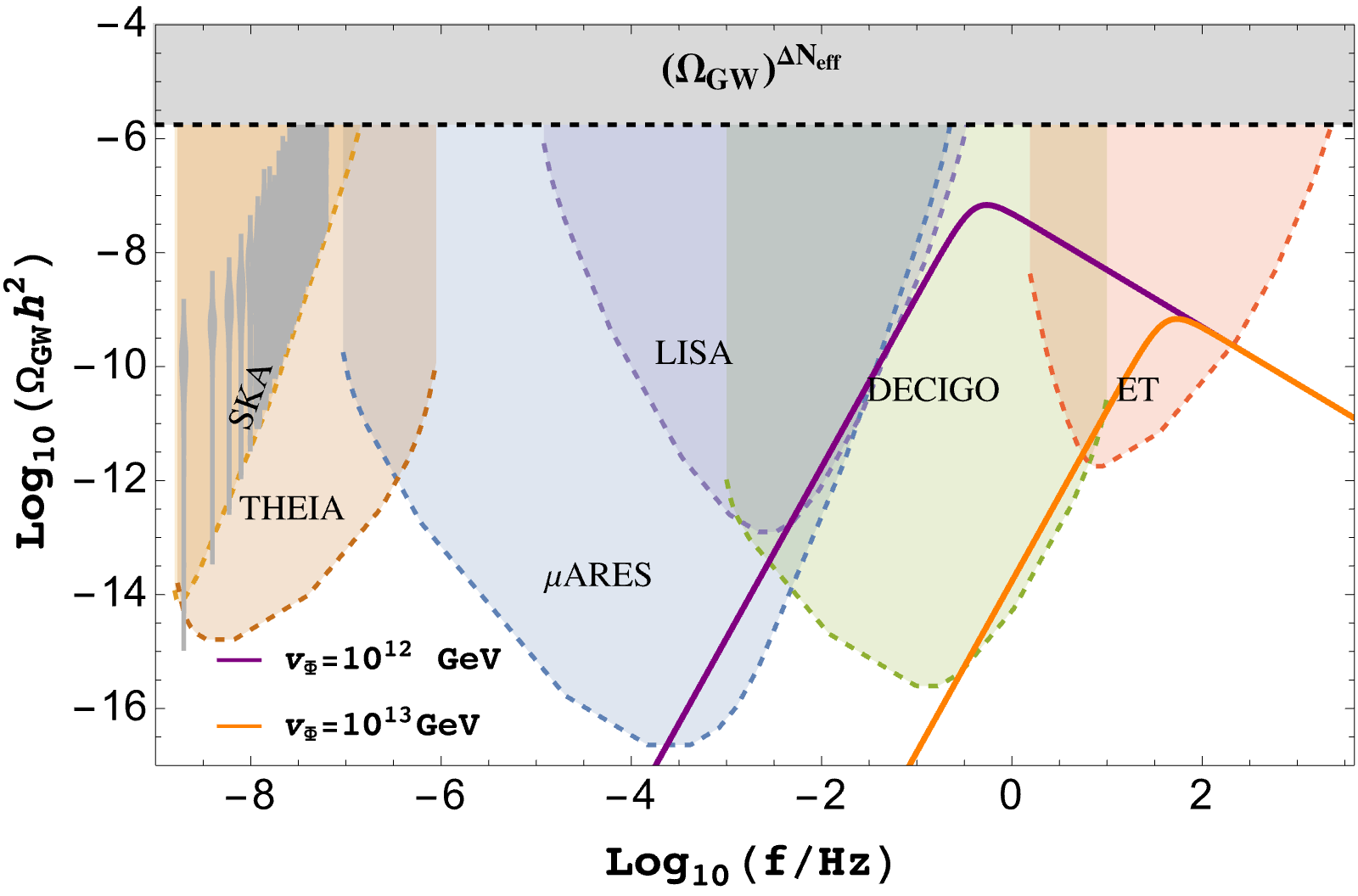}
  \includegraphics[scale=0.4]{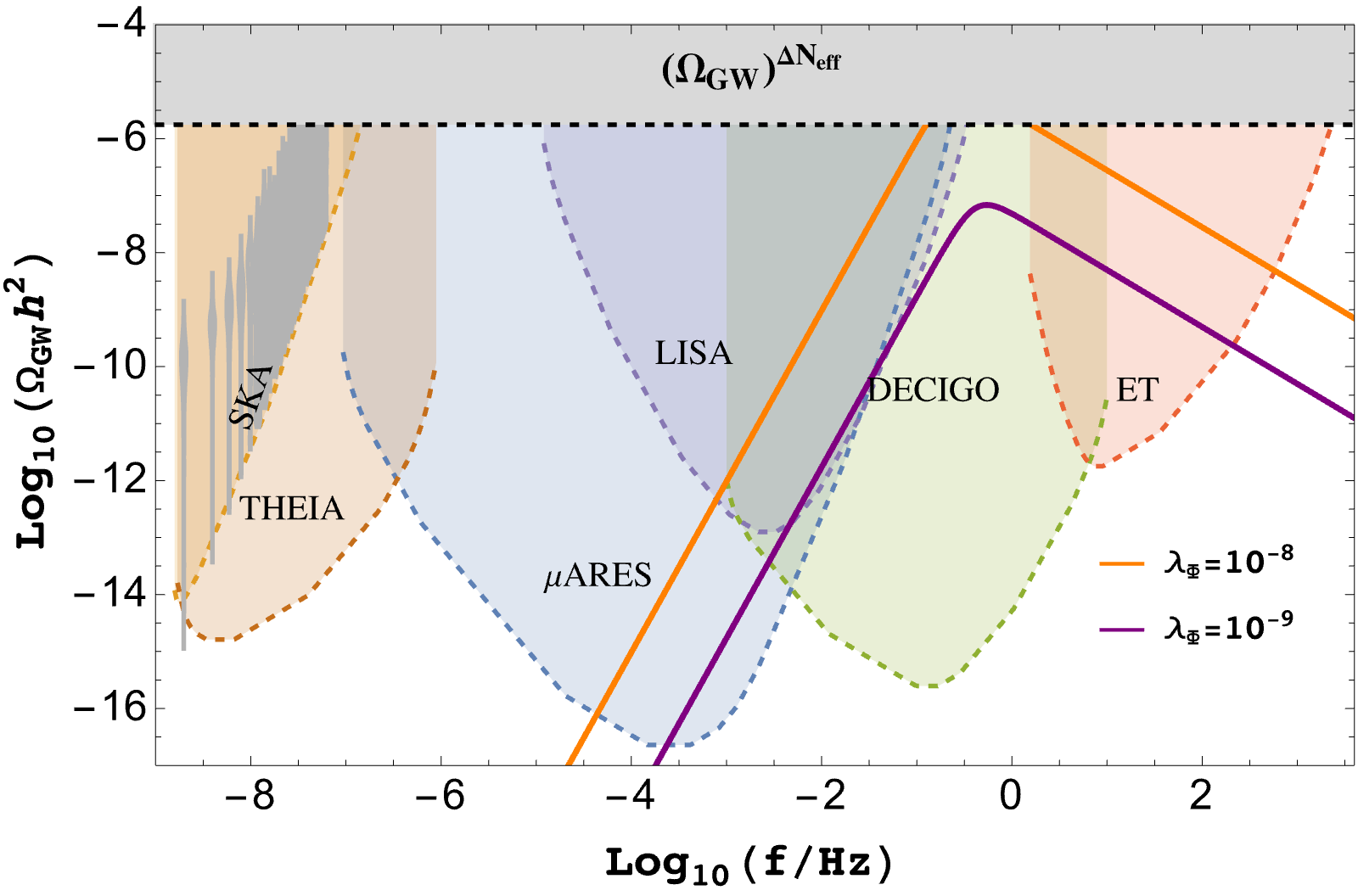} 
\caption{The orange and purple curves show the GW spectrum generated from the annihilation of the DW. In both the panels we set $\Lambda_{\rm QG}=M_{\rm Pl}$ while fixing $\lambda_\Phi=10^{-9
}$ (left panel) and $v_\Phi=10^{12}~\rm{GeV}$ (right panel).  The gray color region denotes the portion excluded from $\Delta N_{\rm eff}$ bound on energy density of GW.}
  \label{fig:DW_spec2}
\end{figure}

As seen from the above discussions, the lower bound on the QG scale from DM and DW requirements remains high because explicit $Z_2$-breaking higher dimensional operators arise at dimension five level in the minimal singlet majoron model.  It is possible to have successful DM and GW phenomenology with a smaller QG scale if higher dimensional operators arise with dimensions more than five. While in the minimal model discussed in this work, we can not prevent dimension five operators suppressed by the QG scale, it is possible with additional gauge symmetries that prevent such operators up to certain dimensions. Similar ideas have already been adopted in the context of the axion quality problem \cite{Barr:1992qq, Dias:2002gg, Carpenter:2009zs}.
Just for the demonstrative purpose, we consider a dimension 7 operator (as it also breaks the $Z_2$ symmetry explicitly) and write the energy bias after replacing field $\Phi$ with its VEV, this leads to a bias contribution to the effective potential of the form
\begin{equation}
    V^{}_{\rm bias} \simeq \frac{v^7_\Phi}{\Lambda^{3}_{\rm QG}}   \; .
    \label{eq:model-bias_dim7}
\end{equation}
 Following this, in Fig. \ref{fig:DW_spec2},  we set $\Lambda_{\rm QG}=M_{\rm Pl}$, where $M_{\rm Pl}$ denotes Planck mass and then we show the GW spectrum for different choices of the model parameters. Now from Eq. \eqref{fpeak}, we find that for the dimension 7 operator, $\Omega_{p}h^2\propto (\lambda_\Phi^2 M_{\rm Pl}^6)/v_\Phi^2$ and $f_p\propto v_\Phi^2/(\lambda_\Phi^{1/4}M^{3/2}_{\rm Pl})$. Clearly, the presence of $M_{\rm Pl}^6$ in numerator provides a huge enhancement in the peak amplitude for the dimension 7 operator. To compensate for such enhancement, a larger value of $v_\Phi$ together with a relatively smaller value of $\lambda_\Phi$ are required. Fig. \ref{fig:DW_spec2} shows that with $v_\Phi\sim 10^{12}$ GeV, $\lambda_\Phi\sim 10^{-9}$ and $\Lambda_{\rm QG}=M_{\rm Pl}$, the GW spectrum generated by the DW annihilation remains within reach of different near future GW experiments. 
 
\begin{figure}[htb!]
  \centering
  \includegraphics[scale=0.51]{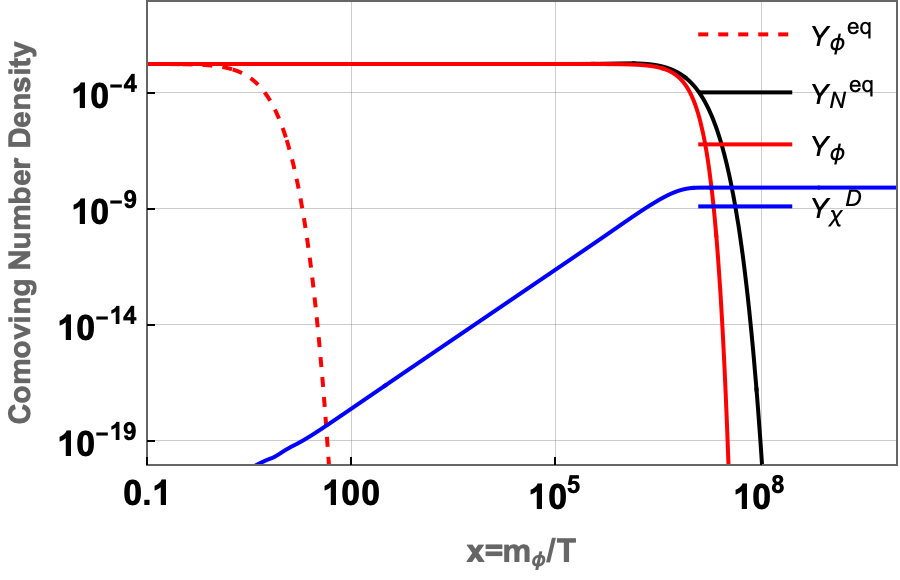}
  \includegraphics[scale=0.31]{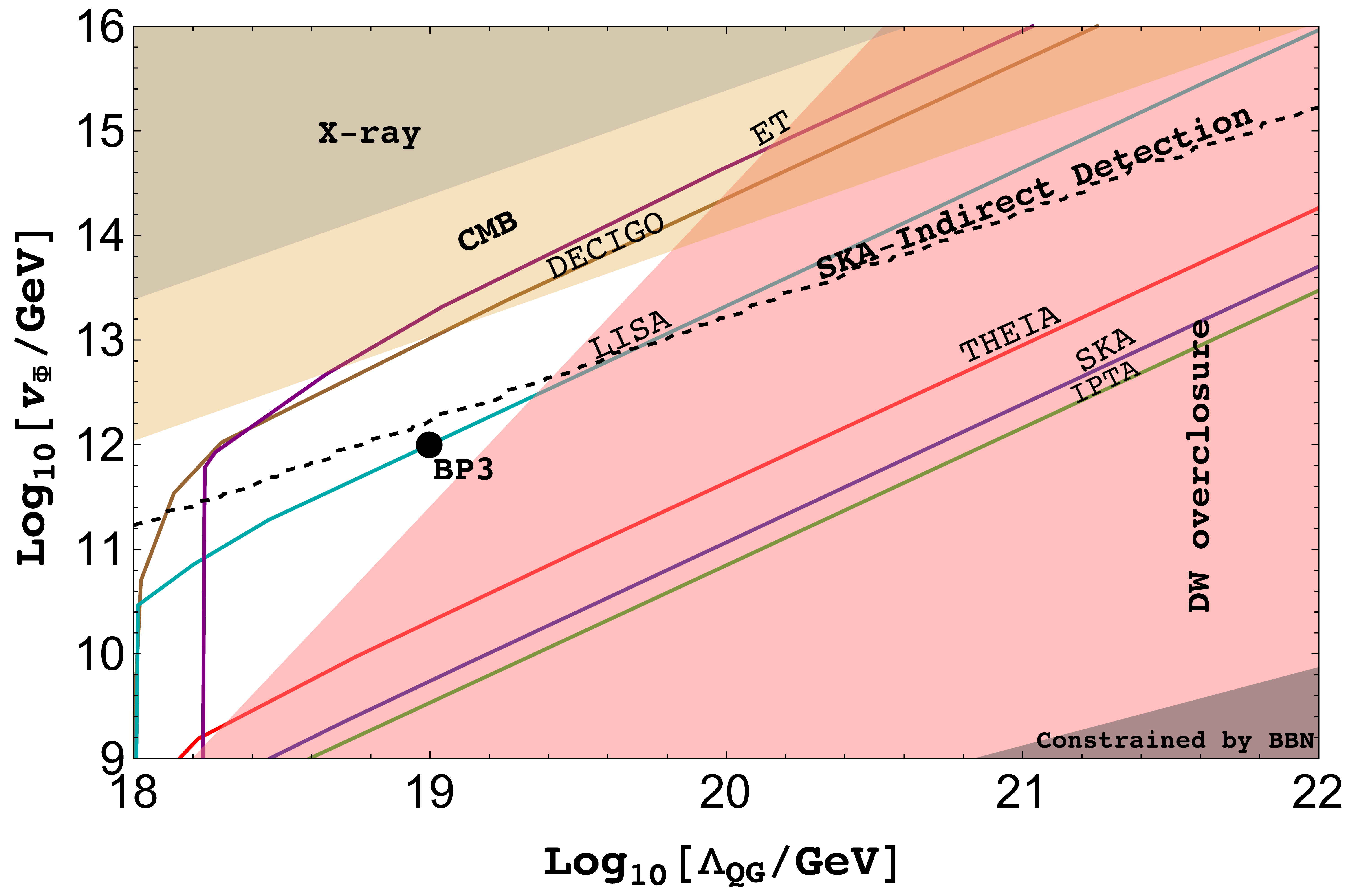}
\caption{Left panel: evolution of comoving number densities for BP3: $v_{\Phi}=10^{12}$ GeV, $\lambda_{\Phi}=10^{-9}$ (which gives $m_{\phi}=4.5\times 10^{7}$ GeV), $m_{N}=20$ GeV and $m_{\chi}=20$ MeV. Right panel: combined constraints on $\Lambda_{\rm QG}$ and $v_\Phi$ with $\lambda_{\Phi}=10^{-9}$. The light gray and yellow-shaded regions denote the excluded regions by X-ray and CMB observations respectively with DM mass 20 MeV. The black dashed line shows the sensitivity of upcoming SKA telescope. The black dot corresponds to BP3. The red, blue, brown, cyan, purple, and green curves represent the sensitivities of THEIA, SKA, DECIGO, LISA, ET and IPTA with SNR = 10. The gray-shaded regions are excluded by the requirement of BBN. The pink-shaded region corresponds to the scenario where DW may overclose the Universe at an early epoch. }
  \label{fig:summary2}
\end{figure}

 Similar to $Z_2$-breaking bias term, DM decay can also occur due to dimension 7 operator of the form
\begin{equation}
    -\mathcal{L}_{\rm decay} \supset \frac{Y_{D_\alpha}}{\Lambda^3_{\rm QG}} \overline{\ell_{L\alpha}}\tilde{H} \chi_R \Phi^3 + {\rm h.c.},
    \label{DMdecay}
\end{equation}
We choose a different benchmark point (BP3) to illustrate the combined result for DM and GW phenomenology.  The chosen values for BP3 are : $v_{\Phi}=10^{12}$ GeV, $\lambda_{\Phi}=10^{-9}$, $m_{N}=20$ GeV, $m_{\chi}=20$ MeV, $\Lambda_{\rm QG}=M_{\rm Pl}$ and majoron mass in sub-eV scale. For BP3 the mass of $\phi$ is found to be $m_{\phi}=4.5\times 10^{7}$ GeV. Although to satisfy correct DM relic, it is sufficient to consider $ 2 m_{N} < m_{\phi}$ as discussed in sub-section \ref{sub1} and \ref{sub2}, $m_{N}$ around TeV scale requires a DM mass around GeV scale (e.g., $m_{N}=10^{4}$ GeV requires $m_{\chi} \sim 1$ GeV). Now, a GeV scale DM mass along with  $v_{\Phi} = 10^{12}$ GeV and $\Lambda_{\rm QG} = M_{\rm Pl}$  results in short DM lifetime in tension with indirect detection limits. Therefore, we choose a smaller $m_{N}=20$ GeV and hence a smaller $m_{\chi}=20$ MeV for BP3. The evolution of relevant number densities together with DM number density for BP3 are shown in the left plot of Fig. \ref{fig:summary2}. The final freeze-in abundance of DM, dominantly produced from decay, is consistent with the observed DM abundance. In the right panel of Fig. \ref{fig:summary2}, we summarise our result for the dimension 7 operator. Here, we notice that the QG scale can be reduced to the Planck scale but only for a relatively large value of $v_\Phi$ and a smaller value of $\lambda_\Phi$. Note that the price we pay to achieve the desired DM and GW phenomenology is a relatively small RHN mass (sub-TeV). As a result, its interaction with majoron becomes very feeble and the majorons never thermalises. This is the reason why we do not see the $\Delta N_{\rm eff}$ bounds from thermalised majoron in Fig. \ref{fig:summary2}. Despite this, the parameter space and BP3 in particular, remain within reach of near-future GW experiments like LISA and DM indirect detection experiments. While light majoron does not give rise to observable $\Delta N_{\rm eff}$ in this scenario, higher dimensional operators can make majoron very heavy due to lower values of QG scale. If majoron becomes heavier than DM, we can have additional freeze-in contribution to DM relic. We find that this contribution is of same order as the contribution from the decay of $\phi$ without changing the generic conclusions. The details of majoron decay contribution to DM relic is given in appendix \ref{appenC}.

\section{Conclusion}
\label{sec5}
We have studied the phenomenology of a dark matter scenario embedded in a global lepton number symmetry $U(1)_L$ focusing on the detection prospects. The spontaneous breaking of global lepton number symmetry generates right-handed neutrino mass or type-I seesaw scale dynamically. Explicit $U(1)_L$ breaking of the quadratic type not only leads to the mass of majoron, the pseudo-NG boson but also reduces the $U(1)_L$ symmetry to a $Z_2$. The effective spontaneous breaking of $Z_2$ symmetry then leads to the formation of domain walls. We consider higher dimensional $Z_2$-breaking operators suppressed by the QG scale as bias necessary for the domain walls to annihilate away while generating stochastic gravitational waves. The same QG scale-suppressed operators also lead to the decay of dark matter. In the minimal singlet majoron model, such QG scale suppressed operators can arise at dimension five level and get tightly constrained from DW, GW as well as DM lifetime criteria. These constraints make the non-thermal production of DM via singlet scalar portal couplings more realistic. We consider the majoron to be in the sub-eV regime and produced thermally. This brings another detection prospect as light majoron can lead to observable $\Delta N_{\rm eff}$ within reach of future CMB experiments. While the scale of QG is required to be $\Lambda_{\rm QG} \gtrsim 10^{25}$ GeV for detectable GW from DW annihilation while satisfying indirect detection bounds on decaying DM, it can be brought down to the Planck scale if explicit global symmetry operators are considered at dimension 7 level instead of dimension 5.

To conclude, the minimal singlet majoron model provides a dynamical origin of the seesaw scale while also accommodating a stable DM without any additional discrete symmetries. The heavy right handed neutrinos not only generate light neutrino masses, but can also lead to successful leptogenesis with resonantly enhanced CP asymmetry. Quantum gravity effects, providing an origin of explicit global symmetry breaking, contribute to the generation of GW from domain wall annihilations while also opening up possible indirect detection aspects due to DM decay. Moreover, along with GW signatures, an observable $\Delta N_{\rm eff}$ keeps the detection prospects of the model very promising while providing a common solution to the neutrino mass, dark matter and baryon asymmetry puzzles.

\acknowledgments
The work of D.B. is supported by the Science and Engineering Research Board (SERB), Government of India grants MTR/2022/000575 and CRG/2022/000603. D.B. also acknowledges the support from the Simons Foundation
(Award Number:1023171-RC) to visit the International Institute of Physics, Natal, Brazil in May 2024 when part of this work was completed. R.R. acknowledges financial support from the STFC Consolidated Grant ST/T000775/1. The work of N.D. is supported by the Ministry of Education, Government of India via the Prime Minister's Research Fellowship (PMRF) December 2021 scheme. N.D. would like to thank Disha Bandyopadhyay for useful discussions. The authors would also like to express special thanks to the organisers of the Workshop in High Energy Physics Phenomenology (WHEPP XVII), IIT Gandhinagar, India where this project was initiated.

\appendix

\section{Coupling of CP-odd and CP-even scalars with $N$ and $\chi$} \label{appenA}

The interaction terms between $\Phi$ and $N_R$ are
\begin{equation}
    \frac{f_{i}}{2} \Phi \overline{N^{c}_{{R}_{i}}} N_{{R}_{i}} + \frac{f_{i}}{2} \Phi^{*} \overline{N_{{R}_{i}}} N^{c}_{{R}_{i}}.
\end{equation}
Writing $N_{i} = N_{{R}_{i}}+N^{c}_{{R}_{i}} = P_{R}N_{i} + P_{L}N_{i}$, we get-
\begin{equation}
    \frac{f_{i}}{2} \Phi \overline{P_{L}N_{i}} P_{R}N_{i} + \frac{f_{i}}{2} \Phi^{*} \overline{P_{R}N_{i}}  P_{L}N_{i}  \nonumber \\
    =\frac{f_{i}}{2} \Phi \overline{N_{i}} P_{R}N_{i} + \frac{f_{i}}{2} \Phi^{*} \overline{N_{i}} P_{L}N_{i} .
\end{equation}
After SSB, we can write $\Phi=\frac{1}{\sqrt{2}}(\phi + v_{\Phi} + i\eta)$. This gives -
\begin{eqnarray}
   &&\frac{f_{i}}{2} \Phi \overline{N_{i}} P_{R}N_{i} + \frac{f_{i}}{2} \Phi^{*} \overline{N_{i}} P_{L}N_{i}  \nonumber \\
    &=& \frac{f_{i}}{2} \frac{1}{\sqrt{2}}(\phi + v_{\Phi} + i\eta) \overline{N_{i}} P_{R}N_{i} + \frac{f_{i}}{2} \frac{1}{\sqrt{2}}(\phi + v_{\Phi} - i\eta) \overline{N_{i}} P_{L}N_{i} \nonumber \\
    &=& \frac{f_{i}}{2\sqrt{2}} \phi\overline{N_{i}}N_{i} + \frac{f_{i}}{2\sqrt{2}} v_{\Phi} \overline{N_{i}}N_{i} + i \frac{f_{i}}{2\sqrt{2}} \eta \overline{N_{i}} \gamma^{5}N_{i}.
\end{eqnarray}
Similarly the interaction terms between $\Phi$ and DM give -
\begin{eqnarray}
    && y_{\chi} \Phi \overline{\chi_{L}} \chi_{R} + y_{\chi} \Phi^{*} \overline{\chi_{R}} \chi_{L} \nonumber \\
    &=& \frac{y_{\chi}}{\sqrt{2}} \phi\overline{\chi}\chi + \frac{y_{\chi}}{\sqrt{2}} v_{\Phi} \overline{\chi}\chi + i \frac{y_{\chi}}{\sqrt{2}} \eta \overline{\chi} \gamma^{5}\chi,
\end{eqnarray}
where $\chi = \chi_{L} + \chi_{R}$.

\section{Cross-section}
\label{appenB}
Here we provide the expressions for relevant cross-sections used in the numerical analysis. 
 
\begin{eqnarray}
 &&(\sigma v)_{N\eta\to N\eta} =\frac{f^4}{2048 \pi  \left(\left(m_{\eta }^2+m_N^2-s\right){}^2-4 m_{\eta }^2 m_N^2\right)} \Biggl (\frac{ (-2 m_N^2 \left(m_{\eta }^2+s\right)+m_N^4+\left(s-m_{\eta }^2\right))}{s^2} \times  \nonumber \\ &&  \Biggl(-2 m_{\eta}^2+\frac{m_{\eta}^4 \left(s-m_N^2\right){}^2 \left(s m_N^2+m_N^4+2 s^2\right)+2 m_{\eta }^{10} \left(m_N^2+s\right)-m_{\eta }^8 \left(m_N^2+s\right) \left(3 m_N^2+s\right)}{\left(s-m_N^2\right){}^2 \left(2 m_{\eta }^6-m_{\eta }^4 \left(3 m_N^2+s\right)+\left(m_N^3-s m_N \right)^2\right)}      \nonumber \\  
&& +m_N^2+5 s \Biggl)  -\frac{2 \left(2 m_{\eta }^4+\left(s-m_N^2\right){}^2\right)}{s-m_N^2} \log \Biggl(\frac{m_{\eta }^4-m_N^2 \left(2 m_{\eta }^2+s\right)+m_N^4}{s \left(2 m_{\eta }^2+m_N^2-s\right)}\Biggl) \Biggl). 
\end{eqnarray}

\begin{eqnarray}
   && (\sigma v)_{\phi\phi\to N\Bar{N}} = -\frac{f^4}{512 \pi  s \left(s-4 m_{\phi}^2\right)} \times  \nonumber\\ && \Biggl(s \sqrt{\frac{\left(s-4 m_{N}^2\right) \left(s-4 m_{\phi}^2\right)}{s^2}} +  \left(8 m_{N}^2-2 m_{\phi}^2+s\right) \log \left(\frac{s \left(\sqrt{\frac{\left(s-4 m_{N}^2\right) \left(s-4 m_{\phi}^2\right)}{s^2}}-1\right)+2 m_{\phi}^2}{2 m_{\phi}^2-s \left(\sqrt{\frac{\left(s-4 m_{N}^2\right) \left(s-4 m_{\phi}^2\right)}{s^2}}+1\right)}\right) \nonumber \\ && + \frac{s \left(16 m_{N}^4+m_{N}^2 \left(s-12 m_{\phi}^2\right)+2 m_{\phi}^4\right) \sqrt{\frac{\left(s-4 m_{N}^2\right) \left(s-4 m_{\phi}^2\right)}{s^2}}}{m_{N}^2 \left(s-4 m_{\phi}^2\right)+m_{\phi}^4} \nonumber \\ && -  \frac{2 \left(-16 m_{N}^4+4 m_{N}^2 s+m_{\phi}^4\right) \log \left(\frac{s \left(\sqrt{\frac{\left(s-4 m_{N}^2\right) \left(s-4 m_{\phi}^2\right)}{s^2}}-1\right)+2 m_{\phi}^2}{2 m_{\phi}^2-s \left(\sqrt{\frac{\left(s-4 m_{N}^2\right) \left(s-4 m_{\phi}^2\right)}{s^2}}+1\right)}\right)}{2 m_{\phi}^2-s} \Biggl).
\end{eqnarray}

\section{Dark matter from Majoron decay}
\label{appenC}
In this work, we have focused on light majorons which remain relativistic during recombination and contribute to $\Delta N_{\rm eff}$. Therefore, we have kept the soft $U(1)_L$ breaking mass term in the sub-eV range. However, higher dimensional operators suppressed by powers of $\Lambda_{\rm QG}$ can push the majoron to higher masses. In the minimal model with dimension five operators suppressed by $\Lambda_{\rm QG}$, the next-to-leading-order (NLO) contribution to majoron mass is $ \sim \sqrt{v^3_\Phi/\Lambda_{\rm QG}}$. For some parts of the parameter space, this contribution can be kept at or below eV scale while for the rest of the parameter space, some tuning will be required between bare mass and NLO mass such that the net majoron mass remains at or below the eV scale. This keeps the $\Delta N_{\rm eff}$ prospects alive as discussed above.

\begin{figure}[htb!]
  \centering
  \includegraphics[scale=1.0]{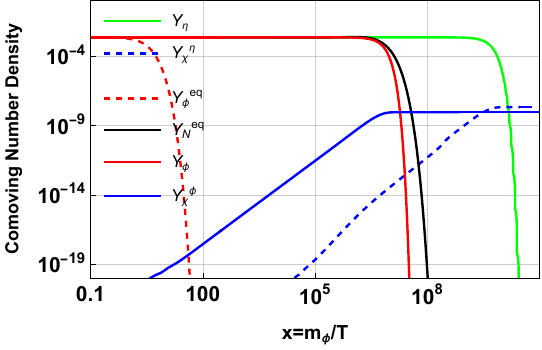}
\caption{Evolution of comoving number densities of different species for BP3.  The solid red and solid green color lines denote the evolution of $\phi$ and $\eta$ respectively. DM abundance from $\phi$ and majoron decay are denoted by solid blue and dashed blue color lines respectively. Both $\phi$ and majoron decay contribute at the same order to the DM abundances as depicted in the figure.}
  \label{fig:majoron_decay}
\end{figure}

On the other hand, for lower values of $\Lambda_{\rm QG}$ as applicable in a scenario where NLO terms arise at dimension seven order, majoron mass $\sim \sqrt{v^5_\Phi/\Lambda^3_{\rm QG}}$ can become substantially large. Since we do not have observable $\Delta N_{\rm eff}$ in the scenario with dimension seven operators as discussed earlier, we do not lose any interesting phenomenology if majorons become heavy and non-relativistic due to NLO operators. However, if majorons become even heavier than DM that is, $m_{\eta} > 2m_{\chi}$, majoron can decay to DM and contribute to the DM abundance. The coupled BEs for majoron and DM can be written as

\begin{eqnarray} \label{BE_majoron}
    \frac{d Y_{\eta}}{dx} &=& - \frac{\beta s}{\mathcal{H}x}\langle \sigma v \rangle _{\eta\eta \to N\Bar{N}} (Y^{2}_{\eta} - (Y^{\rm eq}_{\eta})^{2})  - 
    \frac{\beta}{\mathcal{H}x}\Gamma_{\eta\to N \Bar{N}} \frac{K_{1}(m_{\eta}/T)}{K_{2}(m_{\eta}/T)} (Y_{\eta}-Y^{\rm eq}_{\eta})\\  &&  -\frac{\beta}{\mathcal{H}x}\Gamma_{\eta\to \chi \Bar{\chi}} \frac{K_{1}(m_{\eta}/T)}{K_{2}(m_{\eta}/T)} Y_{\eta}, \nonumber \\
    \frac{d Y_{\chi}}{dx} &=& \frac{\beta}{\mathcal{H}x} \Gamma_{\phi\to \chi \Bar{\chi}}\frac{K_{1}(m_{\phi}/T)}{K_{2}(m_{\phi}/T)} Y_{\phi} + \frac{\beta}{\mathcal{H}x} \Gamma_{\eta\to \chi \Bar{\chi}}\frac{K_{1}(m_{\eta}/T)}{K_{2}(m_{\eta}/T)} Y_{\eta} + \frac{\beta s}{\mathcal{H}x} \langle \sigma v \rangle _{N\Bar{N} \to \chi\Bar{\chi}} (Y^{\rm eq}_{N})^{2}.
\end{eqnarray}
 To compare the production of DM from majoron decay and from $\phi$ decay, we plot the comoving number densities for BP3 in Fig. \ref{fig:majoron_decay}. For BP3, $v_{\Phi}=10^{12}$ GeV, $\lambda_{\Phi}=10^{-9}$ ($m_{\phi}=4.5\times 10^{7}$ GeV), $m_{N}=20$ GeV, $m_{\chi}=20$ MeV, $\Lambda_{\rm QG}=M_{\rm Pl}$. The majoron mass is determined by the dimension $7$ operator which is calculated as $m_{\eta} = \sqrt{\frac{21}{4\sqrt{2}}\frac{v_{\Phi}^5}{\Lambda_{\rm QG}^{3}}}\simeq 61 \text{ GeV}$. Fig. \ref{fig:majoron_decay} shows that the DM yield from $\phi$ and majoron $\eta$ decay by solid blue and dashed blue lines respectively. About equal amount of DM is produced from both the $\phi$ and majoron decay as the the branching ratio of $\phi\to \chi \Bar{\chi}$ and $\eta \to \chi \Bar{\chi}$ are similar. However, the production of DM from majoron decay is delayed compared to $\phi$ decay due to smaller decay width of majoron.

\bibliographystyle{JHEP}
%\bibstyle{apsrev}
\bibliography{ref, ref1, ref2, ref3, refs}

\end{document}